\documentclass[12pt, onecolumn,draftcls]{IEEEtran}
\usepackage{enumerate}
\usepackage{amsmath,amssymb,bm,amsfonts}
\usepackage{graphics,graphicx,epsfig,subfigure}
\usepackage{algorithm,algorithmic}
\usepackage{theorem}
\usepackage{threeparttable}
\usepackage{stfloats}
\usepackage{cite}
\usepackage{multirow}
\usepackage{mathrsfs}
\usepackage{color}
\usepackage{multirow}
\usepackage{subeqnarray}
\usepackage{cases}
\usepackage{wasysym}

\begin{document}

\title{ Distributed Hybrid Power State Estimation under PMU Sampling Phase Errors}
\author{Jian Du, Shaodan Ma,  Yik-Chung Wu and H. Vincent Poor
}

\maketitle
\IEEEpeerreviewmaketitle
\begin{abstract}
 Phasor measurement units (PMUs) have the advantage of providing direct measurements of power states.
However, as the  number of PMUs in a power system is limited, the traditional supervisory control and data acquisition (SCADA) system cannot be
replaced by the PMU-based system overnight.
Therefore, hybrid power state estimation taking advantage of both systems is important.
As experiments show that sampling phase errors among PMUs are inevitable in practical deployment,
this paper proposes a distributed power state
estimation algorithm under PMU phase errors.
The proposed distributed algorithm
only involves local computations and limited information exchange between neighboring areas, thus alleviating the heavy communication burden compared to the centralized approach.  Simulation results show that the performance of the proposed algorithm is very close to that of centralized optimal hybrid state estimates without sampling phase error.
\end{abstract}


\section{Introduction}
Due to the time-varying nature of power generation and consumption,
state estimation in the power grid has always been a fundamental function for real-time monitoring of electric power networks \cite{Monticelli}.
The knowledge of the state vector at each bus, i.e., voltage magnitude and phase angle, enables the energy management system (EMS) to perform various crucial tasks, such as
bad data detection, optimizing power flows, maintaining system stability and reliability \cite{State-SPM}, etc.
Furthermore, accurate state estimation is also the foundation  for the creation and operation of real-time energy markets \cite{Economy}.

In the past several decades, the supervisory control and data acquisition (SCADA) system, which consists of  hardware for signal input/output, communication networks, control equipment, user interface and software \cite{AhmadBook},
has been universally established in the electric power industry, and
installed in virtually all EMSs around the world to manage large and complex power systems.
The large number of remote terminal units (RTUs)   gather  local bus voltage magnitudes, power injection and  current flow magnitudes,  and send them to the master terminal unit to perform centralized state estimation.
As these measurements are nonlinear functions of the power states, the state estimation programs are formulated as iterative reweighted least-squares solution \cite{AburBook,GBGSPM}.

The invention of phasor measurement units (PMUs) \cite{AdventPMU,PMUhistory} has made it possible to measure power states  directly, which is infeasible with SCADA systems.
In the ideal case where PMUs are deployed at every bus, the power state can be simply measured, and this is preferable to the traditional SCADA system.
However, in practice
there are only sporadic  PMUs deployed in the power grid due to expensive installation costs.
In spite of this, through careful placement of PMUs \cite{PMUplace1,PMUplaceGBG,PMUplaceLiXiao}, it is still possible to make the power state observable.
As PMU measurements are linear functions of power states in
rectangular coordinates,
once the observability requirement has been satisfied,
the network state can be obtained by centralized linear least-squares \cite{LS1}.

Despite the advantage of PMUs over SCADA, the traditional SCADA system cannot be replaced by a PMU-based system overnight, as  the SCADA system involves long-term significant investment, and is currently working smoothly in existing power systems.
Consequently, hybrid state estimation with both SCADA and PMU measurements is appealing.
One straightforward methodology is to  simultaneously process both SCADA and PMU raw measurements \cite{XiaoLiPMU}.
However, this simultaneous data processing, which
leads to a totally different set of estimation equations,
 requires significant changes to existing  EMS/SCADA systems \cite{CD, State-SPM}, and is not preferable in practice.
In fact, incorporating PMU measurements with minimal change to the
SCADA system is an important research problem in the power industry \cite{CD}.

In addition to the challenge of integrating PMU with SCADA data, there are also other practical concerns that need to be considered.
Firstly, it is usually assumed that PMUs  provide synchronized sampling of voltage and current signals \cite{ThorpBook} due to the Global Positioning System (GPS) receiver included in the PMU.
However, tests \cite{AsynPMU2} provided by a joint effort between the U.S. Department of Energy and the North American Electric Reliability Corporation
show that PMUs from multiple vendors can yield up to $\pm277.8\mu$s sampling
phase errors (or $\pm6^{\circ}$ phase error in a $60$Hz power system)
due to different delays in the instrument transformers used by different vendors.
Sampling phase mismatch in PMUs
will make the  state estimation problem nonlinear, which offsets the original motivation for introducing PMUs.
It is important to develop state estimation algorithms that are robust to sampling phase errors.

Secondly, with fast sampling rates of PMU devices, a centralized approach, which requires gathering of measurements through propagating
a significant computational
large amounts of data from  peripheral nodes to  a central processing unit,
imposes heavy communication burden across the whole network and imposes a significant computation burden at the  control center.
Decentralizing the computations across different control areas and fusing information in a hierarchical structure or aggregation tree has thus been investigated in \cite{M-state1, M-state2, M-state3, M-state5, M-state6}.
However, these approaches need to meet the requirement of local observability of all the control areas.
Consequently, fully distributed state estimation scalable with network size is preferred \cite{XiaoLiPMU, GBGTPS, XieLeSmartGrid}.

In view of above problems, this paper proposes a distributed power state estimation algorithm, which only involves local computations  and information exchanges with immediate neighborhoods, and  is suitable for implementation in large-scale power grids.
In contrast to \cite{XiaoLiPMU, GBGTPS} and \cite{XieLeSmartGrid}, the proposed distributed algorithm integrates the data from both the SCADA system and PMUs while keeping the existing SCADA system  intact, and the observability problem is bypassed.
The challenging problem of  sampling phase errors  in PMUs is also considered.
Simulation results show that after convergence the proposed algorithm performs  very close to that of the ideal case which assumes
perfect synchronization among PMUs, and centralized information processing.

The rest of this paper is organized as follows.
The state estimation problem  with hybrid SCADA and PMU measurements under sampling errors is presented in Section~\ref{hybrid}.
In Section~\ref{vmpgamma}, a convergence guaranteed  distributed state estimation method is derived.
Simulation results are presented in  Section~\ref{explain}   and
 this work is concluded in Section~\ref{conclusion}.

\textit{Notation}: Boldface uppercase and lowercase letters will be used for matrices and vectors, respectively.
$\mathbb{E}\{\cdot\}$ denotes the expectation of its argument and
$\jmath \triangleq \sqrt{-1}$.
Superscript $T$ denotes transpose.
The symbol $\bm I_N$ represents the $N\times N$ identity matrix.
{The probability density function (pdf) of a random vector $\bm x$ is denoted by $p(\bm x)$, and the conditional pdf of $\bm x$ given $\bm v$ is denoted by $p(\bm x|\bm v)$.}
$\mathcal{N}(\bm x|\bm \mu, \bm R)$ stands for the  pdf of a Gaussian random vector $\bm x$ with mean $\bm \mu$ and covariance matrix $\bm R$.
$\textrm{Bldiag}\{\cdot\}$
denotes the block diagonal concatenation of input arguments.
The symbol $ \propto$ represents a linear scalar relationship between two real-valued functions.
The cardinality of a set $\mathcal{V}$ is denoted by $|\mathcal{V}|$ and the difference between two sets $\mathcal{V}$ and $\mathcal{A}$ is denoted by $\mathcal{V}\setminus \mathcal{A}$.

\section{Hybrid Estimation Problem Formulation}\label{hybrid}
The power grid consists of buses and branches, where a bus can represent a
generator or a load substation, and a branch can stand
for a transmission or distribution line, or even a transformer.
The knowledge of the bus state (i.e., voltage magnitude and phase angle) at each bus enable the power management system to perform functions such as
contingency analysis, automatic generation control, load forecasting and optimal power flow, etc.

Conventionally, the power state is estimated from a set of nonlinear functions with measurements from the SCADA system.
More specifically, a group of RTUs are deployed by the power company at selected buses.
An RTU at a  bus can measure not only injections and voltage magnitudes at the bus but also active and reactive power flows on the branches linked to this bus.
These measurements are then transmitted to the SCADA control center for state estimation.
However, as injections and power flows are nonlinear functions of power states, an iterative method with high complexity is often needed.
The recently invented PMU has the advantage of directly measuring the power states of the bus where it is placed and the current in the branches directly connected
to it.
Through careful PMU placement \cite{Gou}, it is possible to estimate the power states of the whole network with  measurements  from a small number of PMUs.
With measurements from both SCADA and PMUs, it is natural to contemplate obtaining a
better state estimate by combining information from both systems (hybrid estimation).
{In the following, we consider a power network with the set of buses denoted by $\mathcal{B}$ and  the subset of buses with PMU measurements denoted by $\mathcal{P}$.}
\subsection{PMU Measurements with Sampling Errors}\label{PMUmodel}
For a power grid, the continuous voltage on bus $i$ is denoted as
$A_i \cos(2\pi f_c t+ \phi_i)$,
with $A_i$ being the amplitude   and $\phi_i$ being the phase angle in radians.
Ideally, a PMU  provides measurements in rectangular coordinates: $ A_i \cos(\phi_i)$ and $A_i \sin(\phi_i)$.  However,
for reasons of sampling phase error \cite{ThorpBook, AsynPMU2} and measurement error,
the measured voltage at bus $i$ would be  \cite{AsynPMU2}
\begin{equation}\label{vol1}
x_{i}^r = A_i \cos(\theta_i+ \phi_i)  + w_{i,E}^r,
\end{equation}
\begin{equation}\label{vol2}
x_{i}^{\jmath} = A_i \sin(\theta_i+ \phi_i)  + w_{i,E}^{\jmath},
\end{equation}
where $\theta_i$ is the phase error induced by an unknown and random sampling delay, and $w_{i,E}^r$ and $w_{i,E}^{\jmath}$ are the Gaussian measurement noises.
On the other hand,
a PMU also measures the current between neighboring buses.
Let the admittance at the branch $\{i,j\}$ be $g_{ij}+\jmath \cdot b_{ij}$,
the shunt admittance at bus $i$ be $\jmath B_i $,
and the transformer turn ratio from bus $i$ to $j$  be
$\rho_{ij} = |\rho_{ij}|\exp\{\jmath \varphi_{ij}\}$.
Under sampling phase error, the real and imaginary parts of the measured current at bus $i$ are given by \cite{AburBook}
\begin{equation} \label{exactCurrent1}
\begin{split}
y_{ij}^r
=
&\kappa_{ij}^1 A_i \cos(\theta_i+ \phi_i) - \kappa_{ij}^2A_i \sin(\theta_i+ \phi_i)\\
&-\kappa_{ij}^3 A_j \cos(\theta_i+ \phi_j)
+\kappa_{ij}^4 A_j \cos(\theta_i+ \phi_j) + w_{i,I}^r,
\end{split}
\end{equation}
\begin{equation} \label{exactCurrent2}
\begin{split}
y_{ij}^{\jmath}=
& \kappa_{ij}^2 A_i \cos(\theta_i+ \phi_i) + \kappa_{ij}^1 A_i \sin(\theta_i+ \phi_i)\\
&-\kappa_{ij}^4 A_j \cos(\theta_i+ \phi_j)
-\kappa_{ij}^3  A_j \sin(\theta_i+ \phi_j) + w_{i,I}^{\jmath},
\end{split}
\end{equation}
where $\kappa_{ij}^1 \triangleq |\rho_{ij}|^2g_{ij}$,
$\kappa_{ij}^2 \triangleq |\rho_{ij}|^2(b_{ij}+B_i)$,
$\kappa_{ij}^3 \triangleq |\rho_{ij}\rho_{ji}|(\cos\varphi_i^jg_{ij}- \sin\varphi_i^j b_{ij})$,
$\kappa_{ij}^4 \triangleq |\rho_{ij} \rho_{ji}|
(\cos\varphi_i^j b_{ij}+\sin\varphi_i^jg_{ij})$, and
$w_{i,I}^r$ and $w_{i,I}^{\jmath}$ are the corresponding Gaussian measurement errors.

In general, since the phase error $\theta_i$ is small (e.g., the maximum sampling phase error measured by the North American SynchroPhasor Initiative is  $6^{\circ}$ \cite{AsynPMU2}), the
standard approximations  $\sin \theta_i \approx \theta_i $ and $\cos \theta_i \approx 1 $ can be applied to (\ref{vol1}) and (\ref{vol2}), leading to \cite{Pengyang}
\begin{equation}\label{app1}
x_{i}^r \approx E^r_i - E^\jmath_i\theta_i + w_{i,E}^r
\end{equation}
\begin{equation}\label{app2}
x_{i}^\jmath  \approx E^\jmath_i + E^r_i\theta_i + w_{i,E}^\jmath,
\end{equation}
where $E_i^r\triangleq A_i\cos(\phi_i)$ and $E_i^{\jmath}\triangleq A_i\sin(\phi_i)$ denote  the  true power state.
Applying the same approximations to (\ref{exactCurrent1}) and (\ref{exactCurrent2}) yields
\begin{equation}\label{appCurrent1}
\begin{split}
y_{ij}^r
\approx
&\kappa_{ij}^1{E}_i^r - \kappa_{ij}^2{E}^{\jmath}_i
-\kappa_{ij}^3{E}^r_j
+\kappa_{ij}^4{E}^\jmath_j\\
&+ \theta_i
\big\{-\kappa_{ij}^2{E}_i^r-\kappa_{ij}^1{E}^{\jmath}_i
+ \kappa_{ij}^4{E}^r_j
+\kappa_{ij}^3{E}^\jmath_j\big\}  + w_{i,I}^r ,
\end{split}
\end{equation}
\begin{equation}\label{appCurrent2}
\begin{split}
y_{ij}^{\jmath}
\approx
&\kappa_{ij}^2{E}_i^r  + \kappa_{ij}^1{E}^{\jmath}_i
-\kappa_{ij}^4{E}^r_j
-\kappa_{ij}^3{E}^\jmath_j\\
&+ \theta_i
\big\{\kappa_{ij}^1{E}_i^r-\kappa_{ij}^2{E}^{\jmath}_i
-\kappa_{ij}^3 {E}^r_j
+\kappa_{ij}^4{E}^\jmath_j\big\} + w_{i,I}^\jmath.
\end{split}
\end{equation}

We gather all the PMU measurements related to bus $i$ as $\bm z_i = [x^r_{i}, x^\jmath_{i}, y^r_{ij_1}, y^\jmath_{ij_1},\ldots, y^r_{ij_n}, y^\jmath_{ij_n}]^T$
where $j_k$ is the index of bus connected to bus $i$, and arranged in ascending order.
Using (\ref{app1}), (\ref{app2}),  (\ref{appCurrent1}) and (\ref{appCurrent2}), $\bm z_i$ can be expressed in a matrix form as \cite{Pengyang}
\begin{equation} \label{linear}
\bm z_{i}
= \sum_{j\in  \mathcal{M}(i)}  \bm H_{ij}\bm s_j
+
\theta_i\sum_{j\in  \mathcal{M}(i)}   \bm G_{ij}\bm s_j + \bm w_{i},
\end{equation}
where $\bm s_i \triangleq [{E}_i^r, {E}^{\jmath}_i]^T$;
$\mathcal{M}(i)$  is the set of all immediate neighboring buses of bus $i$ and also includes bus $i$;
$\bm H_{ij}$ and $\bm G_{ij}$ are  known matrices containing elements $0$, $1$,
$\kappa_{ij}^1$, $\kappa_{ij}^2$, $\kappa_{ij}^3$ and $\kappa_{ij}^4$;
and the measurement error vector
$\bm w_{i}$ is  assumed to be Gaussian
$\bm w_{i}\sim \mathcal{N}(\bm w_{i}|\bm 0,  \sigma_i^2 \bm I)$,
with $\sigma_i^2 $ being the $i^{th}$ PMU's measurement error variance \cite{PMUvar}.

Gathering all the local measurements $\{\bm  z_i\}_{i \in \mathcal{P}}$ and stacking these observations with increasing order on $i$ as a vector $\bm z$, the system observation model is
\begin{equation}\label{linearPMU}
\bm z = \bm H \bm s + \bm\Theta
\bm G\bm s + \bm w,
\end{equation}
where $\bm s$, $\bm w$ and $\bm \theta$ contain  $\{\bm s_i\}_{i\in \mathcal{B}}$, $\{\bm w_i\}_{i\in \mathcal{P}}$ and $\{\theta_i\}_{i\in \mathcal{P}}$ respectively, in ascending order with respect to $i$;
$\bm\Theta \triangleq   \textrm{Bldiag}\{\theta_i\bm I_{2|\mathcal{M}(i)|}, \ldots, \theta_j\bm I_{2|\mathcal{M}(j)|}\}$ with $i,j\in \mathcal{P}$
 arranged in ascending order;
and $\bm H$ and $\bm G$ are obtained by stacking  $\bm H_{i,j}$ and $\bm G_{i,j}$ respectively, with padding zeros in appropriate locations.
Since $\bm w_i$ is Gaussian, $\bm w$ is also Gaussian with covariance matrix $\bm R = \textrm{Bldiag}\{\sigma_i^2 \bm I_{2|\mathcal{M}(i)|},\ldots, \sigma_j^2 \bm I_{2|\mathcal{M}(j)|}\}$, with $i, j\in \mathcal{P}$, and
the conditional pdf of (\ref{linearPMU}) given $\bm s$ and $\bm\theta$ is
\begin{equation}\label{likelihood1}
p(\bm z|\bm s, \bm \theta ) =
\mathcal{N}(\bm z|(\bm H  +
\bm\Theta\bm G)\bm s,\bm R).
\end{equation}

\subsection{Mixed Measurement from SCADA and PMUs}\label{HybridModel}

For the existing SCADA system, the RTUs  measure
active and reactive power flows in network branches, bus injections and voltage magnitudes at buses.
The  measurements of the whole network by the SCADA system can be described as \cite{State-SPM}
$
\bm \zeta= \bm g(\bm \xi )+ \bm n, $
where $\bm \zeta$  is the vector of the measurements from RTUs in the SCADA system,
$\bm \xi  \triangleq [A_1, \phi_1, A_2,\phi_2, \ldots, A_{|\mathcal{B}|}, \phi_{|\mathcal{B}|}]^T$,
and $\bm n \sim \mathcal{N}(\bm n|\bm 0, \bm W)$
is the measurement noise from RTUs.
Due to the nonlinear function $\bm g(\cdot)$, $\bm \xi $ can be determined by the iterative reweighted least-squares algorithm \cite{EarliestState}, and it was shown in \cite{EarliestState} that with proper initialization,
such a  SCADA-based state estimate $\hat{\bm \xi }$ converges
to the maximum likelihood (ML) solution with
covariance matrix
$\bm \Upsilon = [\nabla \bm g(\bm \xi )^T \bm W^{-1}\nabla \bm g(\bm \xi )  ]^{-1}|_{\bm \xi  = \hat{\bm \xi }}$, where $\nabla \bm g(\bm \xi )$ is the partial derivative of $\bm g$ with respect to $\bm \xi $.

While there are many possible ways of integrating measurements from SCADA and PMUs, in this paper, we adopt the approach that keeps the SCADA system intact, as the SCADA system involves long-term investment and is running smoothing in current power networks.
In order to incorporate the polar coordinate state estimate $\hat{\bm \xi }$ with the PMU measurements in (\ref{linearPMU}),
the work \cite{CD} advocates transforming  $\hat{\bm \xi }$ into rectangular coordinates,
denoted as $\hat{\bm s}_{\textrm{SCADA}}\triangleq\mathcal{T}(\hat{\bm \xi })$.
Due to the invariant property of the ML estimator \cite{Kay}, $\hat{\bm s}_{\textrm{SCADA}}$ is also the ML estimator in rectangular coordinates.
Furthermore, the mean and covariance of $\hat{\bm s}_{\textrm{SCADA}}$ can be approximately computed using the linearization method or unscented transform \cite{SimonFilter}.
For example, based on the linearization method,
Appendix A shows that the mean and covariance matrix of $\hat{\bm s}_{\textrm{SCADA}}$  are $\bm s$ and
$\bm\Gamma_{\textrm{SCADA}}=\nabla \mathcal{T}({\bm \xi } )\bm \Upsilon \nabla [\mathcal{T}({\bm \xi } )]^T|_{\bm \xi  = \hat{\bm \xi }}$, respectively.

{When considering hybrid state estimation, the information from SCADA can be viewed as prior information for the estimation based on PMU measurements.}
{From the definition of minimum mean square error (MMSE) estimation, the optimal estimate of $\bm s$
is given by}
$
\hat{\bm s} \triangleq \int \bm s p(\bm s|\bm z) d\bm s
$,
where $ p(\bm s|\bm z)$ is the posterior distribution.
Since $\bm s$, the unknown vector to be estimated, is coupled with the nuisance parameter $\bm \theta$,
 the posterior distribution of $\bm s$ has to be obtained from $p(\bm s|\bm z)=\int p(\bm \theta,\bm s|\bm z)d\bm \theta$, and we have
\begin{equation}
\hat{\bm s} = \int\int \bm s p(\bm \theta,\bm s|\bm z) d\bm \theta d\bm s. \nonumber
\end{equation}
As $p(\bm \theta,\bm s|\bm z) = \frac{1}{p(\bm z)}p(\bm z|\bm \theta,\bm s)p(\bm s)p(\bm \theta)$ where $p(\bm s)$ and $p(\bm\theta)$ denote the prior distribution of $\bm s$ and $\bm\theta$ respectively,
and $p(\bm z)=\int p(\bm z, \bm \theta, \bm s)d\bm \theta d\bm s=
\int\int p(\bm z|\bm \theta, \bm s)p(\bm \theta) p(\bm s)d\bm \theta d\bm s$ is the normalization constant,
we have the MMSE estimator
\begin{equation}\label{marginal1}
\hat{\bm s} = \frac{\int\int \bm s p(\bm \theta) p(\bm s)p(\bm z|\bm \theta,\bm s) d\bm \theta d\bm s}{\int\int p(\bm \theta) p(\bm s)p(\bm z|\bm \theta, \bm s)d\bm \theta d\bm s}.
\end{equation}
The distributions $p(\bm s)$ and $p(\bm \theta)$ are detailed as follow.
\begin{itemize}
\item
For $p(\bm s)$, it can be obtained from the distribution of the state estimate from the SCADA system.  While $\hat{\bm s}_{\textrm{SCADA}}$  is asymptotically (large data records) Gaussian with mean $\bm s$ and covariance $\bm \Gamma_{\textrm{SCADA}}$ according to the properties of ML estimators, in practice the number of observations in SCADA state estimation is small and the exact distribution of $\hat{\bm s}_{\textrm{SCADA}}$ under finite observation is in general not known.
In order not to incorporate prior information that we do not have,
the maximum-entropy (ME) principle  is adopted.
In particular, given the mean and covariance of $\hat{\bm s}_{\textrm{SCADA}}$,
the maximum-entropy (or least-informative) distribution
is the Gaussian distribution  with the corresponding mean and covariance \cite{Kendalls2,Xia}, i.e., $p(\hat{\bm s}_{\textrm{SCADA}})\approx \mathcal{{N}}(\hat{\bm s}_{\textrm{SCADA}}|\bm s, \bm\Gamma_{\textrm{SCADA}})$.
{According to the Gaussian function property that positions of the mean and variable can be exchanged without changing the value of the Gaussian pdf,
we have}
\begin{equation}\label{sPrior}
 p(\bm s) \approx
\mathcal{{N}}(\bm s|\hat{\bm s}_{\textrm{SCADA}}, \bm\Gamma_{\textrm{SCADA}}).
\end{equation}
\item
For $p(\bm \theta)$, we adopt the truncated Gaussian model:
\begin{equation} \label{TG}
\begin{split}
p(\theta_i)
&=\mathcal{TN}(
\theta_i|\underline{\theta_i}, \bar{\theta}_{i},\tilde{v}_i,\Tilde{C}_i )
\\
&\triangleq
[U(\theta_i - \underline{\theta_i} )-U(\theta_i - \bar{\theta}_{i} )]
\frac{\mathcal{N}(\theta_i|\tilde{v}_i, \Tilde{C}_i)}
     {\textrm{erf}(\frac{\bar{\theta}_{i}-\tilde{v}_i}{\Tilde{C}^{1/2}_i})
     -\textrm{erf}(\frac{\underline{\theta_i}-\tilde{v}_i}{\Tilde{C}^{1/2}_i})},
\end{split}
\end{equation}
where
$\underline{\theta_i}$ and $\bar{\theta}_{i}$ are lower and upper bounds of the truncated Gaussian distribution, respectively;
$U(x)$ is the unit step function, whose value is zero for negative $x$ and one for non-negative $x$;
$\tilde{v}_i$ and $\Tilde{C}_i$ are the mean and covariance of the original,
non-truncated Gaussian distribution;
and $\textrm{erf}(x)\triangleq\frac{1}{\sqrt{2\pi}}\int_0^x \exp\{-\frac{y^2}{2}\}dy$.
Moreover, the first order moment of (\ref{TG}) is
\begin{equation}\label{1stMoment}
\begin{split}
\tilde\varpi_i
=&\mathbb{E}\{\theta_i\}\\
=&
 \tilde{v}_i- \Tilde{C}_i^{1/2}
\frac{ \mathcal{N}(\bar{\theta}_i| \tilde{v}_i, \Tilde{C}_i )
      -\mathcal{N}(\underline{\theta_i}| \tilde{v}_i, \Tilde{C}_i )}
     {\textrm{erf}(\frac{\bar{\theta}_{i}-\tilde{v}_i}{\Tilde{C}^{1/2}_i})
     -\textrm{erf}(\frac{\underline{\theta_i}-\tilde{v}_i}{\Tilde{C}^{1/2}_i})}\\
\triangleq&
 \Xi_1[\underline{\theta_i},\bar{\theta}_{i},
\tilde{v}_i,\Tilde{C}_i]
\end{split}
\end{equation}
and the second order moment is
\begin{equation}\label{2ndMoment}
\begin{split}
\tilde\tau_i
=&\mathbb{E}\{\theta_i^2\}\\
=& \tilde{v}_i^2
 - 2\tilde{v}_i \Tilde{C}_i^{1/2}
\frac{ \mathcal{N}(\bar{\theta}_i| \tilde{v}_i, \tilde{C}_i )
      -\mathcal{N}(\underline{\theta_i}| \tilde{v}_i, \tilde{C}_i )}
     {\textrm{erf}(\frac{\bar{\theta}_{i}-\tilde{v}_i}{\tilde{C}^{1/2}_i})
     -\textrm{erf}(\frac{\underline{\theta_i}-\tilde{v}_i}{\tilde{C}^{1/2}_i})}\\
&+\tilde{C}_i
\Bigg\{1-
\frac{ (\bar{\theta}_i- \tilde{v}_i)
\mathcal{N}(\bar{\theta}_i| \tilde{v}_i, \tilde{C}_i )
      -(\underline{\theta_i}- \tilde{v}_i)\mathcal{N}(\underline{\theta_i}| \tilde{v}_i, \tilde{C}_i )}
     {\tilde{C}_i^{1/2}\big[\textrm{erf}(\frac{\bar{\theta}_{i}-\tilde{v}_i}{\tilde{C}^{1/2}_i})
     -\textrm{erf}(\frac{\underline{\theta_i}-\tilde{v}_i}{\tilde{C}^{1/2}_i})\big]}
     \Bigg\}\\
\triangleq & \Xi_2 [\underline{\theta_i},\bar{\theta}_{i},
\tilde{v}_i,\Tilde{C}_i].
\end{split}
\end{equation}
In general, the parameters of $p(\theta_i)$ can be obtained through pre-deployment measurements.
For example, the truncated range $[\underline{\theta_i}, \bar{\theta}_{i}]$ is founded to be $[-{6\pi}/180,{6\pi}/{180}]$  according to the test results \cite{ AsynPMU2}.
$\tilde{v}_i$ and $\Tilde{C}_i$  can also be obtained from a histogram generated during PMU testing \cite{M-TruncatedEst}.
On the other extreme, (\ref{TG}) also incorporates the case when we have no statistical information about the unknown phase error: setting $[\underline{\theta_i}, \bar{\theta}_{i}]=[-\pi,\pi]$, $\tilde{v}_i=0$, and $\Tilde{C}_i=\infty$, giving an uniform distributed $\theta_i$ in one sampling period of the PMU.
Further, as $\theta_i$
are independent for different $i$, we have
\begin{equation}\label{thetaPrior}
p(\bm \theta)=\prod_{i\in \mathcal{P}} p(\theta_i).
\end{equation}

\end{itemize}

\textsl{Remark 1:}
{Generally speaking, the phase errors in different PMUs may not be independent depending on the synchronization mechanism.
However, tests [16, p. 35] provided by the joint effort between the U.S. Department of Energy and the North American Electric Reliability Corporation
show that the phase errors of PMU measurements is mostly due  to the individual instrument used to obtain the signal from the power system.
Hence it is reasonable to make the assumption that the phase errors in different PMUs are independent.}

\textsl{Remark 2:}
Under the assumption that all the phase errors $\{\theta_i\}_{i\in \mathcal{P}}$ are zero,
(\ref{marginal1}) reduces to
\begin{equation} \label{reducemarginal1}
\hat{\bm s}
= \int \bm s\frac{ p(\bm s) p(\bm z|  \bm s )}{\int p(\bm s)p(\bm z|  \bm s ) d\bm s} d\bm s
.
\end{equation}
Since both $p(\bm s)$ and $p(\bm z|  \bm s )$ are Gaussian, according to the property that the product of Gaussian pdfs is also a Gaussian pdf \cite{Papoulis}, we have that $p(\bm s)p(\bm z|  \bm s )$ is Gaussian.
Moreover, since $\int p(\bm s) p(\bm z|  \bm s ) d\bm s$ is independent of $\bm s$, the computation of $\hat{\bm s}$
in (\ref{reducemarginal1}) is equivalent to maximizing
$p( \bm s) p(\bm z|  \bm s )$ with respective to $\bm s$, which is expressed as
\begin{equation} \label{max}
\begin{split}
\hat{\bm s}
=&\max_{\bm s}p(\bm s) p(\bm z|  \bm s )
 \\
=&\max_{\bm s}
\left\{-||\hat{\bm s}_{\textrm{SCADA}}-\bm s||^2_{\bm\Gamma_{\textrm{SCADA}}}
- ||\bm z- (\bm H  +
\bm\Theta\bm G)\bm s||^2_{\bm R} \right\}.
\end{split}
\end{equation}
Interestingly, (\ref{max}) coincides with the weighted least-squares (WLS) solution in \cite{CD}.

\section{State Estimation under Sampling Phase Error}\label{vmpgamma}
Given all the prior distributions and the likelihood function,
(\ref{marginal1}) can be written as
$\hat{\bm s} = \int\int \bm s   p( \bm \theta, \bm s  |\bm z)  d\bm \theta d\bm s$, where
$ p( \bm \theta, \bm s  |\bm z)\propto
p( \bm \theta)p( \bm s)p(\bm z| \bm \theta, \bm s  )$.
The integration is complicated
as $\theta_i$ is coupled with $\{\bm s_j\}_{j\in \mathcal{M}(i)}$,
and its expression is not analytically tractable.
Furthermore,
the dimensionality of the state space of the integrand
(of the order of number of buses in a power grid, which is typically more than a thousand)
prohibits direct numerical integration.
In this case, approximate schemes need to be resorted to.
One example is the
Markov Chain Monte Carlo (MCMC)
method, which approximates the distributions and integration operations using a large number of random samples \cite{MCMC}.
However, sampling methods can be computationally demanding, often limiting their use to small-scale problems.
Even if it can be successfully applied, the solution is centralized, meaning that
the network still  suffers from heavy communication overhead.
In this section, we  present another approximate method whose distributed implementation can be easily obtained.
\subsection{Variational Inference Framework}\label{A}
The goal of variational inference (VI) is to find a tractable variational distribution $q(\bm \theta, \bm s )$ that closely approximates the true posterior distribution
$p(\bm\theta, \bm s|\bm z) \propto p(\bm \theta) p(\bm s)p(\bm z|\bm \theta, \bm s)$.
The criterion for finding the approximating $q(\bm\theta, \bm s)$ is to minimize the Kullback-Leibler (KL) divergence between $q(\bm\theta, \bm s)$ and
$p(\bm\theta, \bm s|\bm z)$ \cite{BishopVI}:
\begin{equation} \label{VIKL}
 \textrm{KL}\left[q(\bm \theta, \bm s)||p(\bm \theta, \bm s|\bm z)\right]
\triangleq
-\mathbb{E}_{q(\bm \theta,\bm s)}
\left\{\ln\frac{p(\bm \theta, \bm s|\bm z)}{q(\bm \theta,\bm s)}\right\}.
\end{equation}

If  there is no constraint on $q(\bm \theta, \bm s)$, then the
KL divergence vanishes when
$q(\bm \theta, \bm s)=
p(\bm \theta, \bm s|\bm z )$.
However, in this case, we still face the intractable integration in (\ref{marginal1}).
In the VI framework, a common practice is to apply the mean-field approximation
$q (\bm \theta, \bm s) = q(\bm \theta)q(\bm s)$.
Under this mean-field approximation, the optimal  $q(\bm\theta)$ and $q(\bm s)$ that minimize the KL divergence in (\ref{VIKL}) are given by \cite{BishopVI}
\begin{equation} \label{VIupdat2}
\begin{split}
q(\bm \theta)
\propto
\exp \left\{\mathbb{E}_{  q(\bm s)}
 \left\{\ln  p(\bm \theta) p(\bm s)p(\bm z|\bm \theta, \bm s)\right\}\right\}
\end{split}
\end{equation}
\begin{equation} \label{VIupdat3}
\begin{split}
 q(\bm s)
\propto
\exp \left\{\mathbb{E}_{ q(\bm\theta) }
\left\{\ln p(\bm \theta)p(\bm s) p(\bm z|\bm \theta, \bm s)\right\}\right\}.
\end{split}
\end{equation}
Next, we will evaluate the expressions for
$q(\bm \theta)$ and $q(\bm s)$   in (\ref{VIupdat2}) and  (\ref{VIupdat3}), respectively.

\begin{itemize}
\item
\setlength{\itemindent}{1em}
\noindent \textbf{Computation of
$q(\bm\theta)$:}
\end{itemize}

Assume $q(\bm s)$ is known and
$\bm \mu\triangleq\mathbb{E}_{q(\bm s)}\{\bm s\}$ and
$\bm P \triangleq\mathbb{E}_{q(\bm s)}\{(\bm s-\bm \mu)(\bm s-\bm \mu)^T\}$ exist.
Furthermore, let
$\bm \mu_i = [\bm \mu]_{2i-1:2i}$ be  the local mean state vector of the
$i^{th}$ bus;
and $\bm P_{i,j} = [\bm P]_{2i-1:2i,2j-1:2j}$ be the local covariance of state vectors between the $i^{th}$ and $j^{th}$ buses.

By substituting the prior distributions $p(\bm s)$ from (\ref{sPrior}), $p(\bm \theta)$ from (\ref{TG})   and the likelihood function from
  (\ref{likelihood1})  into (\ref{VIupdat2}), the variational distribution $q(\bm \theta)$ is shown in Appendix B to be
\begin{equation}\label{be-theta2}
\begin{split}
q(\bm \theta)
\propto
\prod_{i\in \mathcal{P}}\mathcal{TN}(\theta_i|\underline{\theta_i},\bar{\theta}_{i},
{v}_i, {C}_i),
\end{split}
\end{equation}
with
\begin{equation}\label{truncatedC}
{C}_i
= \frac{\tilde{C}_i }{\sigma_i^{-2}\textrm{Tr}\{
{\bm A}_{i,2}\}\tilde{C}_i+ 1 },
\end{equation}
\begin{equation}\label{truncatedv}
{v}_i
={C}_i\big[\tilde{v}_i/\tilde{C}_i+ \sigma_i^{-2}
\textrm{Tr}\big\{\bm z_i\sum_{j\in  \mathcal{M}(i)}
(\bm G_{ij}{\bm \mu}_j)^T
-{\bm A}_{i,1}
\big\}\big],
\end{equation}
where ${\bm A}_{i,1}
=\sum_{j\in  \mathcal{M}(i)}\bm H_{ij}\big({\bm P}_{j,j} + {\bm \mu}_j {\bm\mu}_j^T\big)\bm G^T_{ij}
+
\sum_{j,k\in  \mathcal{M}(i), j\neq k}\bm H_{ij}\big({\bm P}_{j,k} + {\bm \mu}_j {\bm\mu}_k^T\big)\bm G^T_{ik} $
and
${\bm A}_{i,2}
=
\sum_{j\in  \mathcal{M}(i)}\bm G_{ij}\big({\bm P}_{j,j} + {\bm \mu}_j {\bm\mu}_j^T\big)\bm G^T_{ij}
+
\sum_{j,k\in  \mathcal{M}(i), j\neq k}\bm G_{ij}\big({\bm P}_{j,k} + {\bm \mu}_j {\bm\mu}_k^T\big)\bm G^T_{ik} $.
Furthermore, the first and second order moments of $q(\bm\theta)$ in (\ref{be-theta2}) can be
easily shown to be $\bm \varpi=[{\varpi}_i \ldots {\varpi}_j]^T$ and
$\textbf{\rm T}=\bm \varpi\bm \varpi^T+\textrm{diag}\{\tau_i -\varpi_i^2 \ldots \tau_j-\varpi_j^2\}
$ respectively, with $i, j\in \mathcal{B}$ and ${\varpi}_i$ and $\tau_i$ computed
according to (\ref{1stMoment}) and (\ref{2ndMoment}) as
\begin{equation}\label{updatetheta1}
{\varpi}_i
= \Xi_1[\underline{\theta_i},\bar{\theta}_{i},
{v}_i,{C}_i],
\end{equation}
\begin{equation}\label{updatetheta2}
{\tau}_{i}
=\Xi_2[\underline{\theta_i},\bar{\theta}_{i},
{v}_i,{C}_i].
\end{equation}
\begin{itemize}
\item
\setlength{\itemindent}{1em}
\noindent \textbf{Computation of
 $q(\bm s)$:}
\end{itemize}

Assume $q(\bm \theta)$ is known and
$\bm\varpi\triangleq\mathbb{E}_{q(\bm \theta)}\{\bm\theta\}$  and
$\textbf{\rm T}\triangleq\mathbb{E}_{q(\bm \theta)}\{\bm\theta\bm\theta^T\}$ exist.
Furthermore, let $\varpi_i = [\bm\varpi]_i$
 be  the local mean of the phase error at the
$i^{th}$ bus,
and $\tau_{i} = [\textbf{\rm T} ]_{i,i}$ be the local second order moment.
By substituting the prior distribution
$p(\bm \theta)$ from
(\ref{thetaPrior}), and the
likelihood function  from (\ref{likelihood1}) into (\ref{VIupdat3}), and performing integration over $\bm \theta$ as shown in Appendix B,
we obtain
\begin{equation}\label{be-s2}
\begin{split}
q(\bm s)
\propto
\mathcal{N}\big(\bm s|{\bm \mu} , {\bm P}\big)
\end{split}
\end{equation}
with the  mean ${\bm \mu}$ and covariance $\bm P$ given by
\begin{equation}\label{s-meanupdated}
\begin{split}
{\bm \mu}
=&
\bm \Gamma_{\textrm{SCADA}}^{-1} \hat{\bm s}_{\textrm{SCADA}} + \bm\Upsilon(\bm H  +  {\bm \Omega} \bm G)^T
\big[(\bm H  +  {\bm \Omega} \bm G){\bm\Upsilon} (\bm H  +  {\bm \Omega} \bm G)^T + \bm R \big]^{-1}\\
&\times[\bm z -(\bm H  +  {\bm \Omega} \bm G)
\bm \Gamma_{\textrm{SCADA}}^{-1} \hat{\bm s}_{\textrm{SCADA}}],
\end{split}
\end{equation}
\begin{equation}\label{s-Covupdated}
{\bm P}
={\bm\Upsilon}-{\bm\Upsilon}(\bm H  +  {\bm \Omega} \bm G)^T
\big[(\bm H  +  {\bm \Omega} \bm G){\bm\Upsilon} (\bm H  +  {\bm \Omega} \bm G)^T + \bm R \big]^{-1}
(\bm H  +  {\bm \Omega} \bm G){\bm\Upsilon},
\end{equation}
respectively, where
${\bm\Upsilon} = [\bm \Gamma_{\textrm{SCADA}}^{-1} + (\bm G^T
({\bm \Lambda}- {\bm \Omega}^2)\bm R^{-1}
\bm G)^{-1}]^{-1}$.

From the expressions for   $q(\bm \theta)$ and  $q(\bm s)$ in
(\ref{be-theta2}) and (\ref{be-s2}), it should be noticed that
 these two functions are coupled.
Consequently, they should be updated iteratively.
Fortunately,  $q(\bm \theta)$ and  $q(\bm s)$
keep the same forms as their  prior distributions, and therefore, only the parameters of each function are involved in the iterative updating.

In summary, let the initial variational distribution  $q^{(0)}(\bm s)$ equal
$p(\bm s)$ in (\ref{sPrior}), which is Gaussian with mean
${\bm \mu} =\hat{\bm s}_{\textrm{SCADA}}$
and covariance matrix ${\bm P}=\bm \Gamma_{\textrm{SCADA}}$.
We can obtain the updated $q^{(1)}(\bm \theta)$ following
(\ref{be-theta2}).
After that,
$q^{(1)}(\bm s)$ will be obtained according to
(\ref{be-s2}).
The process is repeated until ${\bm \mu} $ converges or a predefined  maximum
number of iterations is reached.
Once the converged $q(\bm\theta)$ and $q(\bm s)$   are obtained, $p(\bm \theta, \bm s|\bm z)$ is replaced by $q(\bm\theta)q(\bm s)$ in
(\ref{marginal1}), and it can be readily shown that $\hat{\bm s}$ equals $\mathbb{E}\{q(\bm s)\}=\bm\mu$.

\subsection{Distributed Estimation}\label{B}
For a large-scale power grid, to alleviate the communication burden on the network
and computation complexity at the control center, it is advantageous to decompose the state estimation algorithm
into computations that are local to each area of the power system and require only limited message exchanges  among immediate neighbors.
From (\ref{be-theta2})-(\ref{truncatedv}), it is clear that
$q(\bm\theta)$ is a product of a number of truncated Gaussian distributions $b(\theta_i)\triangleq  \mathcal{TN}(\theta_i|\underline{\theta_i},\bar{\theta}_{i},
{v}_i, {C}_i)$, with each component
 involving measurements only from bus $i$ and parameters relating bus $i$ and its immediate neighboring buses.
Thus the estimation of $\theta_i$ can be performed locally  at each bus.

However, this is not true for $\bm s_i$ in (\ref{be-s2}).
To achieve   distributed computation for the power state $\bm s_i$,
a mean-field approximation is applied to $q(\bm s)$, and we write
$q(\bm s)=\prod_{i\in \mathcal{B}} b(\bm s_i)$.
Then, the variational distribution is in the form
$\prod_{i\in \mathcal{P}}b(\theta_i)\prod_{i\in \mathcal{B}}b(\bm s_i)$.
Since the goal is to derive a distributed algorithm,
it is also assumed that each bus  has access only to the   mean and variance of its own state from SCADA estimates, i.e.,
$p(\bm s)
\thickapprox \prod_{i\in \mathcal{B}} p(\bm s_i)
= \prod_{i\in \mathcal{B}}\mathcal{N}(\bm s_i|\bm \gamma_i, \bm \Gamma_i )$,
with $\bm \gamma_i = [\hat{\bm s}_{\textrm{SCADA}}]_{2i-1:2i}$ and
$\bm \Gamma_i = [\bm P_{\textrm{SCADA}}]_{2i-1:2i,2i-1:2i}$.
Then, the optimal variational distributions $b(\bm s_i)$ and
$b(\theta_i)$ can be obtained through minimizing the following KL divergence:
\begin{equation}\label{KL2}
\begin{split}
&\textrm{KL}\bigg\{\prod_{i\in \mathcal{P}}b(\theta_i)\prod_{i\in \mathcal{B}}b(\bm s_i)
\big|\big|
\frac{
p(\bm z|\bm \theta, \bm s)
\prod_{i\in \mathcal{P}}p(\theta_i)
\prod_{i\in \mathcal{B}}p(\bm s_i)
}
{\int\int p(\bm z|\bm \theta, \bm s)\prod_{i\in \mathcal{B}}p(\bm s_i)
\prod_{i\in \mathcal{P}}p(\theta_i) d\{\theta_i\}_{i\in \mathcal{P}}d\{\bm s_i\}_{i\in \mathcal{B}}}
\bigg\}\\
&
\propto
-
\mathbb{E}_{\prod_{j\in \mathcal{P}}b(\theta_j)\prod_{j\in \mathcal{B}}b(\bm s_j)  }
\bigg\{\ln
\frac{p(\bm z|\bm \theta, \bm s)\prod_{i\in \mathcal{P}}p(\theta_i)
\prod_{i\in \mathcal{B}}p(\bm s_i)
}
{\prod_{i\in \mathcal{P}}b(\theta_i)\prod_{i\in \mathcal{B}}b(\bm s_i)}\bigg\}.
\end{split}
\end{equation}
Similarly to (\ref{VIupdat2}) and (\ref{VIupdat3}), the
$b(\theta_i)$ and $b(\bm s_i)$ that minimize (\ref{KL2}) are given by
\begin{equation} \label{VIupdat22}
\begin{split}
b(\theta_i)
\propto
\exp \Big\{\mathbb{E}_{\prod_{j\in \mathcal{P}\setminus i}b(\theta_j)
\prod_{i\in \mathcal{B}}p(\bm s_i)  }
\big\{ \ln p(\bm z|\bm \theta, \bm s)\prod_{i\in \mathcal{P}}p(\theta_i)
\prod_{i\in \mathcal{B}}p(\bm s_i)
\big\} \Big\}
\quad i\in \mathcal{P},
\end{split}
\end{equation}
\begin{equation} \label{VIupdat333}
\begin{split}
b(\bm s_i)
\propto
\exp \Big\{\mathbb{E}_{\prod_{i\in \mathcal{P}}b(\theta_i)
\prod_{j\in \mathcal{B}\setminus i}b(\bm s_j)}
\big\{ \ln p(\bm z|\bm \theta, \bm s)\prod_{i\in \mathcal{P}}p(\theta_i)
\prod_{i\in \mathcal{B}}p(\bm s_i)
\big\}\Big\}
\quad i\in \mathcal{B}.
\end{split}
\end{equation}
Next, we will evaluate the expressions for
$b(\theta_i)$ and $b(\bm s_i)$   in (\ref{VIupdat22}) and  (\ref{VIupdat333}), respectively.
\begin{itemize}
\item
\setlength{\itemindent}{1em}
\noindent \textbf{Computation of
$b(\theta_i)$:}
\end{itemize}

Assume $b(\bm s_i)$ is known for all $i\in \mathcal{B}$ with
mean and covariance denoted by $\bm \mu_i$ and $\bm P_{i,i}$, respectively.
The $b(\theta_i)$ in (\ref{VIupdat22}) can be obtained from
$q(\bm \theta)$ in (\ref{be-theta2})
by setting $\bm P_{i,j}=0$ if $i\neq j$, and we have
\begin{equation}\label{btheta}
\begin{split}
b(\theta_i)
\propto
\mathcal{TN}(\theta_i|\underline{\theta_i},\bar{\theta}_{i},
{v}_i, {C}_i)
\end{split}
\end{equation}
with
\begin{equation}\label{vbCC}
{C}_i
= \frac{\tilde{C}_i }{\sigma_i^{-2}\textrm{Tr}\{
{\bm B}_{i,2}\}\tilde{C}_i+ 1 }
\end{equation}
\begin{equation} \label{vbtheta}
{v}_i
={C}_i\big[\tilde{v}_i/\tilde{C}_i+ \sigma_i^{-2}
\textrm{Tr}\big\{\bm z_i\sum_{j\in  \mathcal{M}(i)}
(\bm G_{ij}{\bm \mu}_j)^T
-{\bm B}_{i,1}
\big\}\big].
\end{equation}
with ${\bm B}_{i,1}
=\sum_{j\in  \mathcal{M}(i)}\bm H_{ij}\big({\bm P}_{j,j} + {\bm \mu}_j {\bm\mu}_j^T\big)\bm G^T_{ij}
+
\sum_{j,k\in  \mathcal{M}(i), j\neq k}\bm H_{ij} {\bm \mu}_j {\bm\mu}_k^T\bm G^T_{ik} $
and
${\bm B}_{i,2}
=
\sum_{j\in  \mathcal{M}(i)}\bm G_{ij}\big({\bm P}_{j,j} + {\bm \mu}_j {\bm\mu}_j^T\big)\bm G^T_{ij}
+
\sum_{j,k\in  \mathcal{M}(i), j\neq k}\bm G_{ij} {\bm \mu}_j {\bm\mu}_k^T\bm G^T_{ik} $.
With ${C}_i$ and ${v}_i$ in (\ref{vbCC}) and (\ref{vbtheta}), to facilitate the computation of $b(\bm s_i)$ in the next step, the first and second order moments of $b(\theta_i)$ are computed through (\ref{1stMoment}) and (\ref{2ndMoment}) as
\begin{equation}\label{updatetheta11}
{\varpi}_{i}
= \Xi_1[\underline{\theta_i},\bar{\theta}_{i},
{v}_i,{C}_i],
\end{equation}
\begin{equation}\label{updatetheta22}
{\tau}_{i}
=\Xi_2[\underline{\theta_i},\bar{\theta}_{i},
{v}_i,{C}_i].
\end{equation}

\begin{itemize}
\item
\setlength{\itemindent}{1em}
\noindent \textbf{Computation of
$b(\bm s_i)$:}
\end{itemize}

Assume $b(\theta_i)$ for all $i\in \mathcal{P}$
are known with first and second order moments denoted by $\varpi_i$ and $\tau_i$, respectively.
Furthermore, it is assumed that $b(\bm s_j)$ for $j \in \mathcal{B}\setminus i$ are also known with their covariance matrices given by $\bm P_{j,j}$.
Now, rewrite
(\ref{VIupdat333}) as
\begin{equation} \label{VIupdat33}
\begin{split}
b(\bm s_i)
\propto
p(\bm s_i)\prod_{j\in \mathcal{M}(i)}
\underbrace{\exp \big\{\mathbb{E}_{b(\theta_j)\prod_{k\in \mathcal{M}(j)\setminus i}b(\bm s_k)}
\{
  \ln p(\bm z_j|\theta_j, \{\bm s_{\tilde{k}}\}_{\tilde{k}\in \mathcal{M}(j)})\}\big\}}_{\triangleq m_{j\rightarrow i}(\bm s_i)}.
\end{split}
\end{equation}
As shown in Appendix C, $m_{j\rightarrow i}(\bm s_i)$ is in Gaussian form
\begin{equation}\label{vmpmessage}
m_{j\rightarrow i}(\bm s_i)\propto \mathcal{N}(\bm s_i|\bm v_{j\rightarrow i}, \bm C_{j\rightarrow i})
\end{equation}
with
\begin{equation}\label{vmp1}
\bm C_{j\rightarrow i}=\sigma_j^2[\bm H_{ji}^T\bm H_{ji} + \varpi_j(\bm G_{ji}^T\bm H_{ji}+\bm H_{ji}^T\bm G_{ji})+ \tau_{j}\bm G_{ji}^T\bm G_{ji}]^{-1},
\end{equation}
\begin{equation}\label{vmp2}
\begin{split}
\bm v_{j\rightarrow i}
&=\sigma_j^{-2}\bm C_{j\rightarrow i}\bigg\{
(\bm H_{ji}+ \varpi_j\bm G_{ji})^T\bm z_j\\
&
-
\sum_{k\in \mathcal{M}(j)\setminus i}
\big[\bm H_{ji}^T\bm H_{jk}+ \varpi_j(\bm G_{ji}^T\bm H_{jk}+ \bm H_{ji}^T\bm G_{jk})+
\tau_j\bm G_{ji}^T\bm G_{jk}  \big]^T\bm \mu_k   \bigg\}.
\end{split}
\end{equation}
Then, putting $p(\bm s_i)
= \mathcal{N}(\bm s_i|\bm \gamma_i, \bm \Gamma_i )$ and (\ref{vmpmessage}) into (\ref{VIupdat33}), we obtain
\begin{eqnarray} \label{appbs}
b(\bm s_i)
&\propto&\mathcal{N}(\bm s_i|\bm \gamma_i, \bm \Gamma_i )\mathcal{N}(\bm s_i|\bm v_{j\rightarrow i}, \bm C_{j\rightarrow i}) \nonumber \\
&\propto& \mathcal{N}(\bm s_i|\bm \mu_i, \bm P_{i,i}),
\end{eqnarray}
with
\begin{equation}\label{sp}
\bm P_{i,i} = (\bm\Gamma_i^{-1} + \sum_{j\in\mathcal{M}(i)} \bm C_{j\rightarrow i}^{-1})^{-1}
\end{equation}
\begin{equation}\label{sm}
\bm \mu_i = \bm P_{i,i}(\bm\Gamma_i^{-1}\bm \gamma_i + \sum_{j\in \mathcal{M}(i)} \bm C_{j\rightarrow i}^{-1}\bm v_{j\rightarrow i}).
\end{equation}

Inspection of (\ref{vmp1}) and (\ref{vmp2}) reveals that these expressions can be readily computed at bus
$j$ and then $\bm C_{j\rightarrow i}$ and $\bm v_{j\rightarrow i}$ can be sent to its immediate neighbouring bus $i$ for computation of $b(\bm s_i) $ according to (\ref{appbs}).

\begin{itemize}
\item
\setlength{\itemindent}{1em}
\noindent \textbf{Updating Schedule and Summary:}
\end{itemize}

From the expressions for $b(\theta_i)$ and $b(\bm s_i)$ in (\ref{btheta}) and (\ref{appbs}), it should be noticed that  these functions are coupled.
Consequently, $b(\theta_i)$ and $b(\bm s_i)$  should be iteratively updated.
Since updating any $b(\theta_i)$ or $b(\bm s_i)$ corresponds to minimizing the KL divergence in (\ref{KL2}),
\begin{algorithm}
\caption{Distributed states estimation}
\label{Alg11}
\begin{algorithmic}[1]
\STATE Initialization:
$\bm \mu_i = [\hat{\bm s}_{\textrm{SCADA}}]_{2i-1:2i}$ and
$\bm P_{i,i}=[\bm P_{\textrm{SCADA}}]_{2i-1:2i;2i-1:2i}$.\\
Neighboring buses exchange $\bm \mu_i$ and $\bm P_{i,i}$.\\
Buses with PMUs update ${\varpi}_{i}$ and ${\tau}_{i}$ via (\ref{updatetheta11}) and (\ref{updatetheta22}).\\
Every bus $i$ computes $\bm C_{i\rightarrow j}$  $\bm v_{i\rightarrow j}$
$\bm P_{i,i}$, $\bm \mu_i$  via (\ref{vmp1}) (\ref{vmp2}) (\ref{sp}) (\ref{sm}),
and sends these four entities to bus $j$, where $j\in \mathcal{M}(i)$.

\FOR{the $l^{th}$ iteration}
\STATE Select a group of buses with the same color.\\
\STATE Buses with PMUs in the group compute ${\varpi}_{i}$ and ${\tau}_{i}$ via (\ref{updatetheta11}) and (\ref{updatetheta22}).\\
\STATE Every bus in the group updates its $\bm C_{i\rightarrow j}$  $\bm v_{i\rightarrow j}$
$\bm P_{i,i}$, $\bm \mu_i$  via (\ref{vmp1}) (\ref{vmp2}) (\ref{sp}) (\ref{sm}), and
sends them out to its neighbor $j$.\\
\STATE Bus $j$ computes
$\bm v_{j\rightarrow k}$ via (\ref{vmp2}) and send to its neighbor $k\in \mathcal{M}(j)$.
\ENDFOR
\end{algorithmic}
\end{algorithm}
the iterative algorithm is guaranteed to converge monotonically to at least a stationary point \cite{BishopVI} and there is no requirement that $b(\theta_i)$ or $b(\bm s_i)$ should be updated in any particular order.
Besides, the variational distributions  $b(\theta_i)$ and $b(\bm s_i)$   in (\ref{btheta}) and (\ref{appbs})
keep the form of truncated Gaussian and Gaussian distributions during the iterations, thus only their parameters are required to be updated.

However, the successive update scheduling might take too long
in large-scale networks.
Fortunately, from (\ref{vmp1})-(\ref{sm}), it is found that
updating $b(\bm s_i)$ only involves  information within two hops   from bus $i$.
Besides, from (\ref{vbCC}) and  (\ref{vbtheta}), it is observed that updating $b( \theta_i)$ only involves information from  direct neighbours of
 bus $i$.
{Since KL divergence is a convex function with respect to each of the factors $b(\bm s_i)$ and $b(\theta_i)$,} if buses within two hops from each other do not update
their variational distributions $b(\cdot)$
at the same time, the KL divergence in (\ref{KL2}) is guaranteed to be decreased in each iteration and the distributed algorithm keeps the monotonic convergence property.  This can be achieved by grouping
the buses using a distance-2
coloring scheme \cite{Color}, which colors all the buses under the principle that buses within a two-hop neighborhood are assigned   different colors and
the number of colors used is the least
(for the IEEE-$300$  system, only $13$ different colors are needed).
Then, all  buses  with the same
color update
at the same time and buses with different colors are updated in succession.
{Notice that the complexity order of the distance-2 coloring scheme is
 $\mathcal Q(\lambda|\mathcal{B}|)$\cite{Color}, where  $\lambda$ is the maximum number of branches linked to any bus.
Since $\lambda$ is usually small compared to the network size
(e.g., $\lambda=9$ for the IEEE 118-bus system),
the
complexity of distance-2 coloring  depends only on the network size and it is independent of the specific topology of the power network.}

In summary, all the buses are first colored by the distance-$2$ coloring scheme, and
the iterative procedure is formally given in Algorithm \ref{Alg11}.
{Notice that
although the  modelling and formulation of state estimation under phase error is complicated, the final result and processing are  simple.
During each iteration, the first and second order moments of the phase error estimate are computed via (\ref{updatetheta11}) and (\ref{updatetheta22});
while the covariance and  mean of the state estimate are computed using (\ref{sp}) and (\ref{sm}).
Due to the fact that computing these quantities at one bus depends on information from neighboring buses, these equations are computed iteratively.}
After convergence, the state estimate is given by $\bm \mu_i$ at each bus.

Although the proposed distributed algorithm advocates each bus to perform computations and message exchanges, but it is also applicable if computations of several buses are executed by a local control center.  Then any two control centers only need to exchange the messages for their shared power states.

\section{Simulation Results and  Discussions}\label{explain}
This section provides results on the numerical tests of the
 developed centralized and distributed state estimators in Section \ref{vmpgamma}.
The network parameters
$g_{ij}$, $b_{ij}$,  $ B_i $,
 $\rho_{ij} $
are loaded from the test cases in MATPOWER$4.0$ \cite{MatPower}.
{ In each simulation, the value at each load bus is varied by adding a uniformly distributed random value within} $\pm10\%$ {of the value in the test case.
Then the power flow program is run to determine the true states.}
The RTUs measurements are composed of active/reactive power injection, active/reactive power flow, and bus voltage magnitude at each bus, which are also generated from MATPOWER$4.0$ and perturbed by independent
zero-mean Gaussian measurement errors
with standard deviation
$1\times 10^{-2}$ \cite{PMUvar}.
For the SCADA system, the estimates $\hat{\bm \xi }$ and $\bm \Upsilon $  are obtained through the classical iterative reweighted least-squares  with
initialization
$[A_i, \phi_i]^T=[1, 0]^T$ \cite{AburBook}.
In general, the proposed algorithms are applicable regardless of the number of PMUs and their placements.
But for the simulation study,
the placement of PMUs is obtained through the method proposed in \cite{Gou}.
As experiments in \cite{AsynPMU2} show the maximum
phase error is $6^{\circ}$ in a 60Hz power system, $\theta_i$ is generated uniformly from
$[-{6\pi}/180,{6\pi}/{180}]$ { for each Monte-Carlo simulation run}.
The PMU measurement errors follow a zero-mean Gaussian distribution with
standard deviation $\sigma_i = 1\times 10^{-2}$ \cite{PMUvar}.
$1000$ Monte-Carlo simulation runs are averaged for each point in the figures.
Furthermore, it is  assumed that bad data from RTUs and PMU measurements has been  successfully handled \cite{CD}, \cite[Chap 7]{PoorGridBook}.

For comparison, we consider the following three existing methods:
1) Centralized WLS \cite{CD} assuming no sampling phase errors in the PMUs.
{Without sampling phase error, (10) reduces to $\bm z=\bm H\bm s + \bm w$.
For this linear model, WLS can be directly applied to estimate $\bm s$.}
{This algorithm serves as
a benchmark for the proposed algorithms.}
2) Centralized WLS under sampling phase errors in the PMUs.
This will show how much degradation one would have if phase errors are ignored.
3) {The centralized alternating minimization (AM) scheme \cite{Pengyang} with} $p(\bm s)$ and $p(\theta_i)$ in (\ref{sPrior}) and (\ref{TG}) incorporated  as prior information.
{In particular, the posterior distribution is maximized alternatively with respect to $\bm s$ and $\bm \theta$.
While updating one variable vector, all others should be kept at the last estimation values.}

\begin{figure}[H]
\centering
\epsfig{file=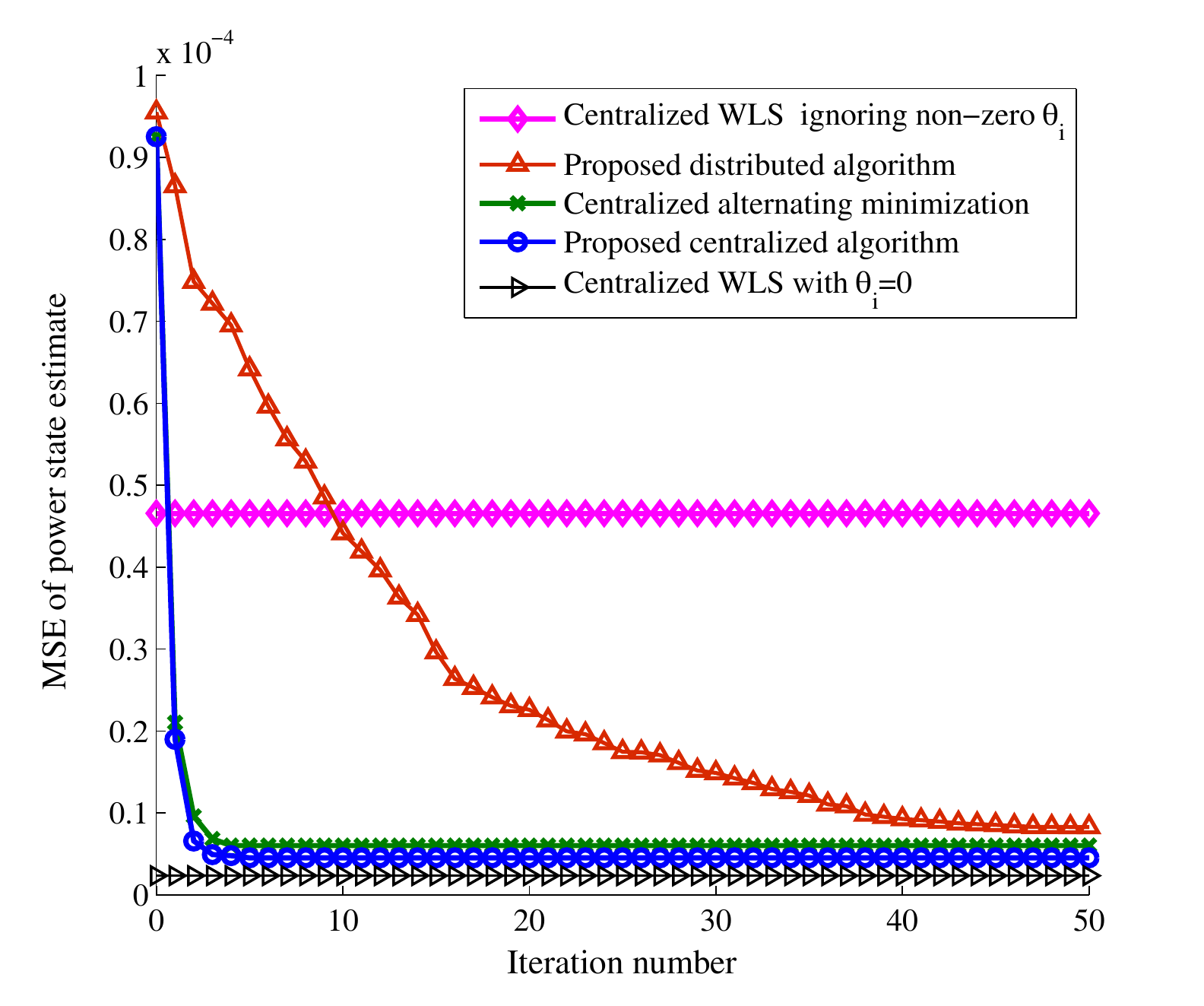, width=3.6in}
\caption{MSE of the power state versus iteration number for the IEEE $118$-bus system.}
\label{State-iter}
\end{figure}

\begin{figure}[H]
\centering
\epsfig{file=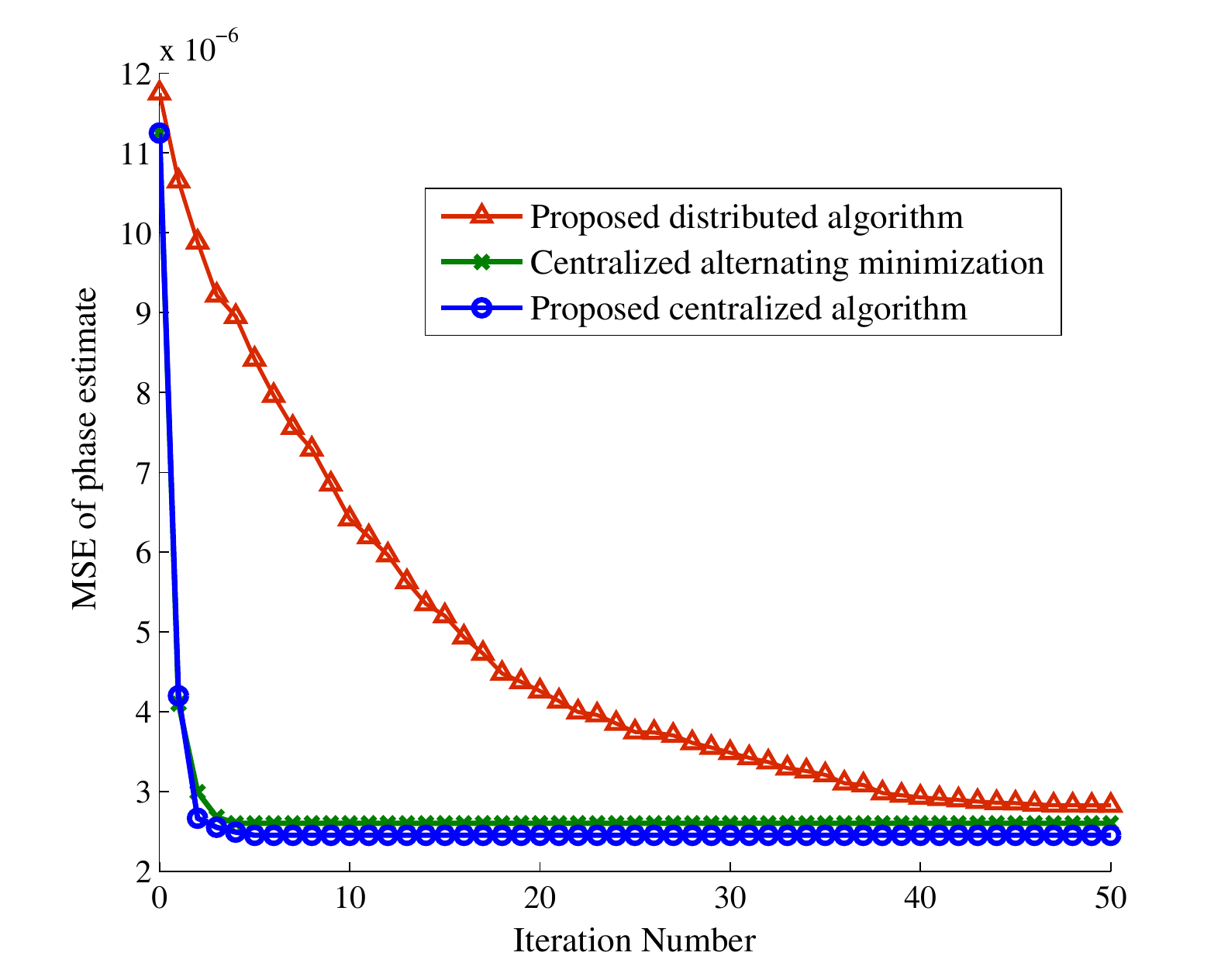, width=3.6in}
\caption{MSE of the phase error versus iteration number for the IEEE $118$-bus system.}
\label{Phase-iter}
\end{figure}

Fig. \ref{State-iter} shows the convergence behavior of the  proposed algorithms with average mean square error (MSE) defined as
$\frac{1}{2|\mathcal{B}|}\sum_{i\in \mathcal{B}}||\hat{\bm s}_i -\bm s_i||^2$.
It can be seen that:
a) The centralized VI approach converges very rapidly and after convergence the corresponding MSE are  very close to the benchmark performance provided by WLS with no sampling phase offset.
Centralized AM is also close to optimal after convergence.
b) The proposed distributed algorithm can also approach the optimal performance after convergence.
The seemingly slow convergence is
a result of sequential updating of
buses with different colors  to guarantee convergence.
If one iteration is defined as one round of updating of all buses, the distributed algorithm would converge only in a few iterations.
On the other hand, the small degradation from the centralized VI solution is due to the fact that in the distributed algorithm, the covariance of states $\bm s_i$ and $\bm s_j$ in prior distributions and variational distributions cannot be taken into account.
c) If the sampling phase error is ignored, we can see that the performance of  centralized WLS  shows significant degradation, illustrating the importance of simultaneous power state and phase error estimation.
Fig. \ref{Phase-iter} shows the MSE of the  sampling phase error estimation
$\frac{1}{|\mathcal{P}|}\sum_{i\in \mathcal{P}}||{\varpi}^{\ast}_i - \theta_i||^2$, where ${\varpi}^{\ast}_i$ is the converged ${\varpi}_i$ in (\ref{updatetheta11}).
It can be seen from the figure that same conclusions as in Fig. \ref{State-iter} can be drawn.

\begin{figure}[H]
\centering
\epsfig{file=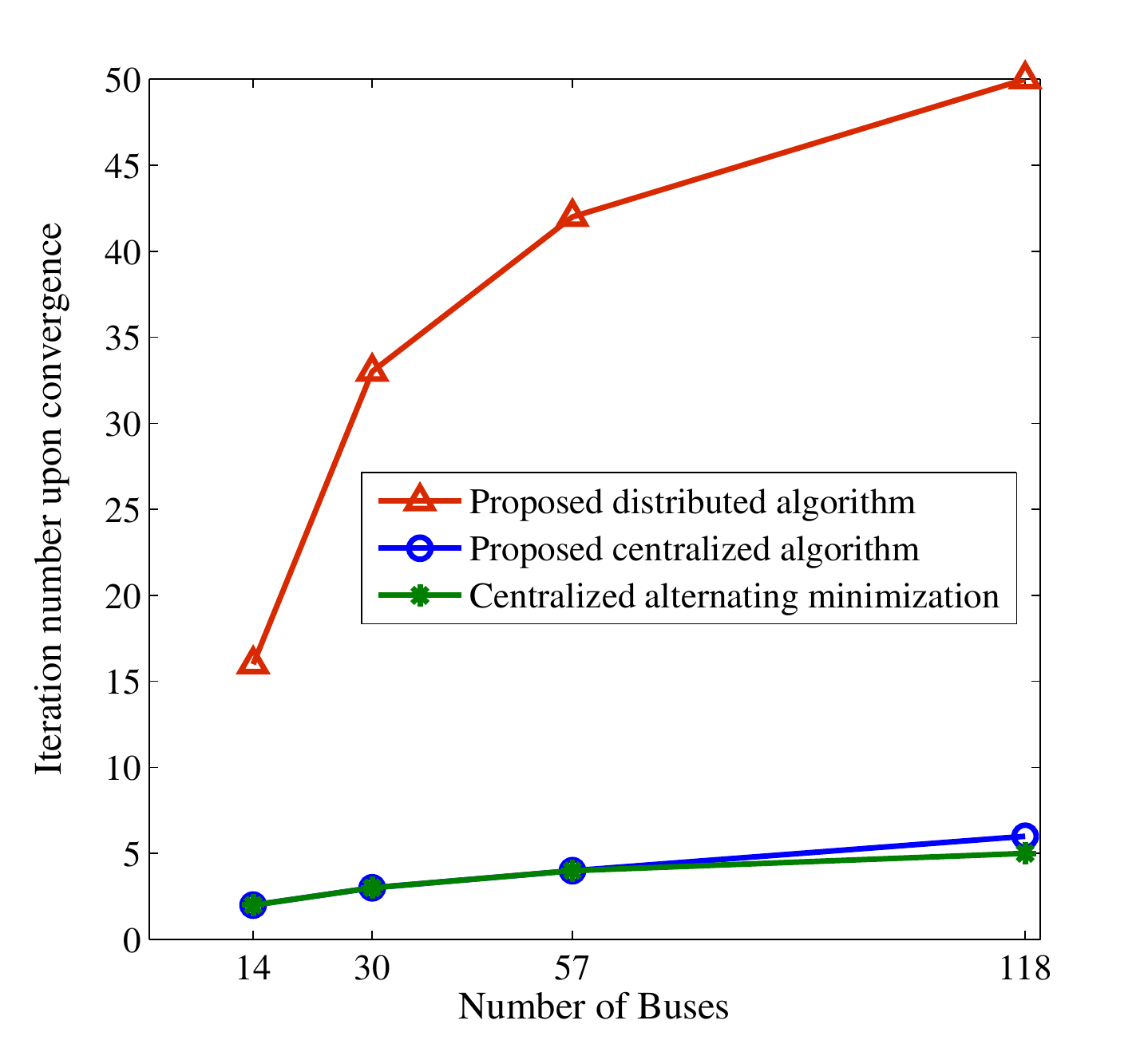, width=3.6in}
\caption{Iteration numbers upon convergence versus the network size.}
\label{Iter-busNo}
\end{figure}

Fig. \ref{Iter-busNo} shows the relationship between iteration number upon convergence versus  the network size.
The seemingly slow convergence of the proposed distributed algorithm is again due to the sequential updating of buses with different colors.
However, more iterations in the proposed distributed algorithm do not mean a larger computational complexity.
In particular, let us consider a network with $|\mathcal B|$ buses.
In the centralized AM algorithm \cite{Pengyang}, for each iteration, the computation for power state estimation is dominated by a $2|\mathcal B|\times 2|\mathcal B|$ matrix inverse and the complexity is   $\mathcal O((2|\mathcal B|)^3)$,
while the computation for phase error estimation is dominated by a $|\mathcal B|\times |\mathcal B|$ matrix inverse and the complexity is   $\mathcal O((|\mathcal B|)^3)$.
Hence, for centralized AM algorithm, in each iteration, the computational complexity order is $\mathcal O(9|\mathcal B|^3)$.
On the other hand, in the proposed distributed algorithm,  the computational complexity of each iteration at each bus is dominated by   matrix inverses with dimension $2$ ((\ref{vmp1}), (\ref{vmp2}) (\ref{sp}) and (\ref{sm})), hence the computational complexity is of order $ \mathcal O(2^3)$, and the complexity of the whole network in each iteration is of order
$\mathcal O(2^3\times |\mathcal B|)$, which is only linear with respect to number of buses.
It is obvious that a significant complexity saving is obtained compared to
the centralized AM algorithm ($\mathcal O(9|\mathcal B|^3)$).
Thus, although the proposed distributed algorithm requires more iterations to converge, the total computational complexity is still much lower than that of its centralized counterpart.
Such merit is important for power networks with high data throughput.

\begin{figure}[H]
\centering
\epsfig{file=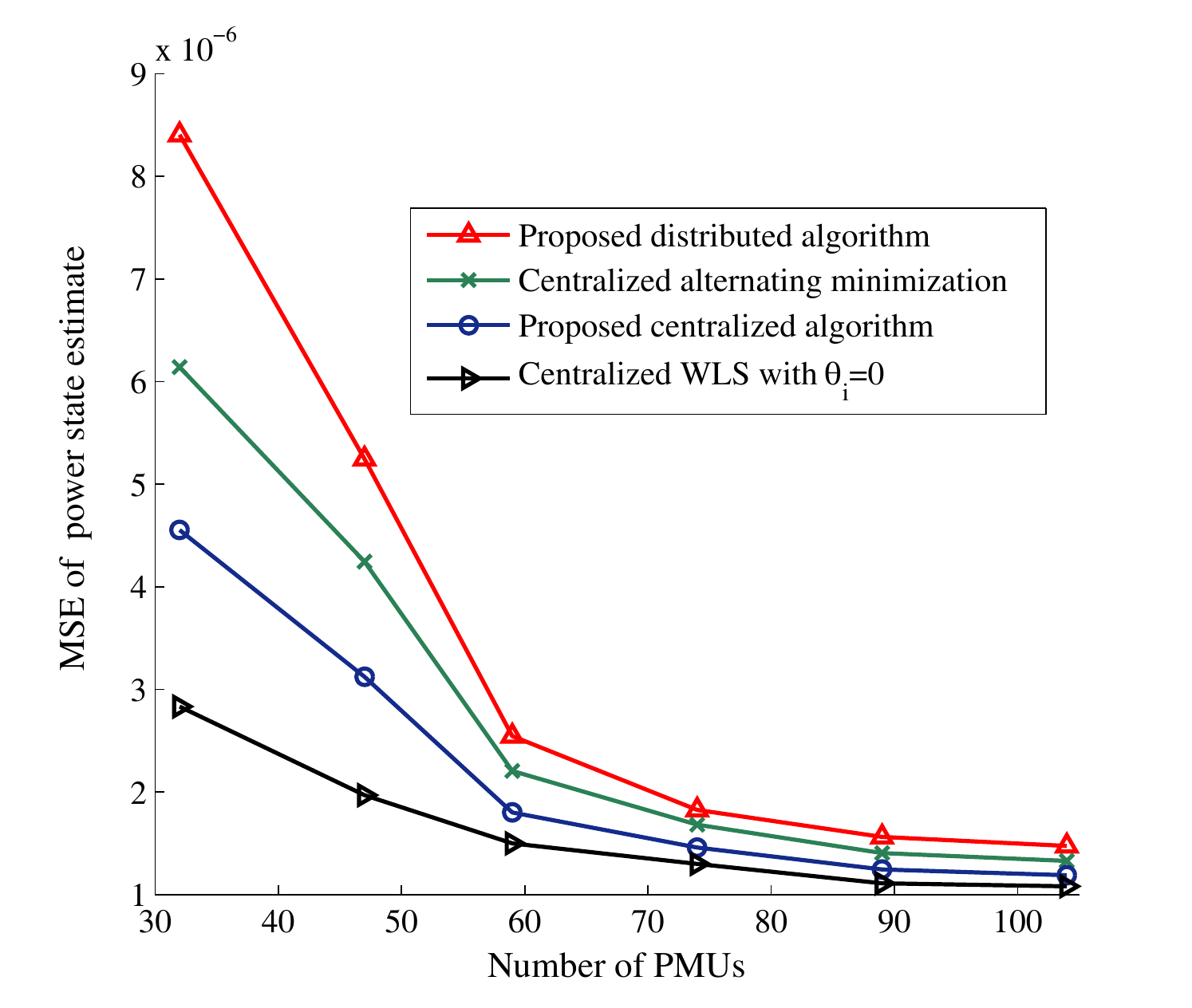, width=3.6in}
\caption{Effect of increasing the number of PMUs  on the power state estimate.}
\label{State-PMUno}
\end{figure}

The effect of using different numbers of PMUs in the IEEE $118$-bus
system is shown in Fig. \ref{State-PMUno}.
First, $32$ PMUs are placed over the network according to \cite{Gou} for full topological observation.
The remaining PMUs, if available, are randomly placed to provide additional measurements. The MSE of state estimation  is plotted versus the number of PMUs.
It is clear that  increasing the number of PMUs is beneficial to  hybrid state estimation.
But the improvement shows diminishing return as the number of PMUs increases.
The curves in this figure allow system designers to choose a tradeoff between estimation accuracy and the number of PMUs being deployed.

\begin{figure}[H]
\centering
\epsfig{file=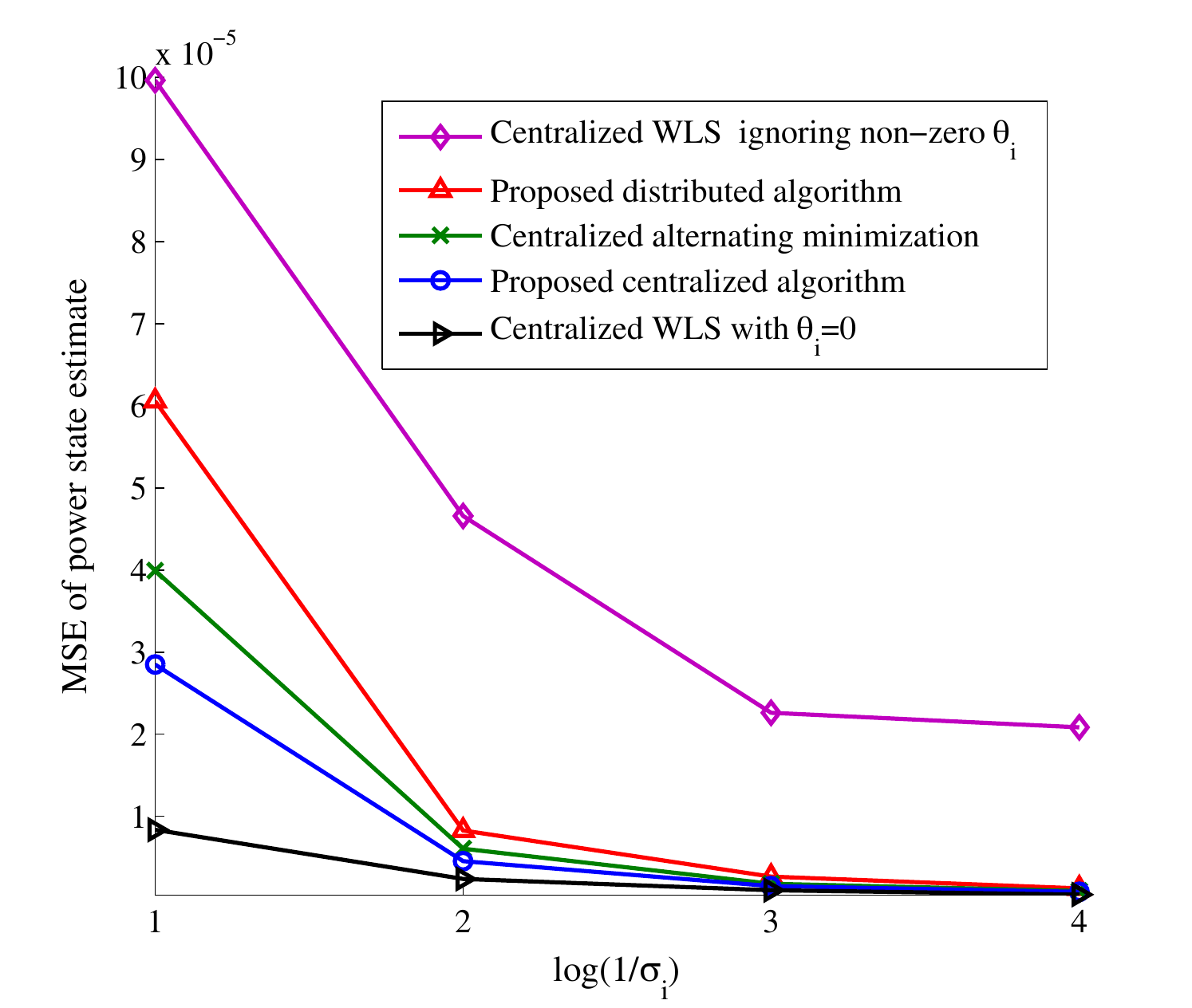, width=3.6in}
\caption{ MSE of power state versus $\log (1/\sigma_i)$, where $\sigma_i$ is the standard deviation of the $i^{th}$ PMU's measurement error.}
\label{MSE-SNR}
\end{figure}

Finally, Fig. \ref{MSE-SNR} shows the MSE versus PMU measurement error variance for the IEEE 118-bus system.
It can be seen that with smaller measurement error variance, the MSE of the proposed distributed method becomes very close to   the optimal performance.
However, if we ignore the sampling phase errors, the estimation MSE
shows a constant gap from that of optimal performance
even if the  measurement error variance tends to zero.
This is because in this case, the non-zero sampling phase dominates the error in the PMU measurements.

\section{Conclusions} \label{conclusion}
In this paper,  a distributed state estimation scheme integrating measurements from a traditional SCADA system and newly deployed PMUs  has been proposed, with the aim that the existing SCADA system is kept intact.
Unknown sampling phase errors among PMUs have  been
incorporated in the
estimation procedure.
The  proposed distributed power state estimation algorithm
only involves limited message exchanges between neighboring buses and is guaranteed to converge.
Numerical results have shown that   the converged state estimates of the distributed algorithm
are very close to those of the optimal centralized  estimates assuming no sampling phase error.

\appendices
\section{ }\label{sect:appendixA}
Let the nonlinear transformation  from polar to rectangular coordinate be denoted by $\mathcal{T}(\cdot)$.
Assuming $\bm \hat{s}_{\textrm{SCADA}} = \mathcal{T}(\hat{\bm \xi  })$ and
performing the first-order Taylor series expansion of $\mathcal{T}(\hat{\bm \xi })$ about the true state $\bm \xi $ yields
\begin{equation}\label{Taylor}
\bm \hat{s}_{\textrm{SCADA}}
= \mathcal{T}(\hat{\bm \xi  })
=\mathcal{T}({\bm \xi } + \Delta\bm \xi   )
\approx\mathcal{T}({\bm \xi }  ) + \nabla \mathcal{T}({\tilde{\bm \xi }} )|_{\tilde{\bm \xi }=\bm \xi }\Delta\bm \xi   ,
\end{equation}
where $\Delta\bm \xi $ is the estimation error from the SCADA system,
and
$\nabla \mathcal{T}(\bm \xi  )$ is the first order derivative of $ \mathcal{T}(\cdot )$, which is a  block diagonal matrix with the $i^{\textrm{th}}$ block
$[\nabla \mathcal{T}(\tilde{\bm \xi })]_{i,i} = \left[\begin{array}{cc}
\cos {\tilde{\theta}}_i & -{E}_i\sin {\tilde{\theta}}_i  \\
\sin {\tilde{\theta}}_i & {E}_i\cos {\tilde{\theta}}_i  \\
\end{array} \right]$ for $i=1,\ldots, M $.
Taking expectation on both sides of (\ref{Taylor}), we obtain
\begin{equation}\label{Taylor-mean}
\mathbb{E}\{\hat{\bm s}_{\textrm{SCADA}} \} \thickapprox \bm s.
\end{equation}
Furthermore, the covariance is
\begin{equation}\label{Taylor-cov}
\bm \Upsilon_{\textrm{SCADA}}
\approx
\nabla \mathcal{T}( {\tilde{\bm \xi }} )\bm \Upsilon \nabla [\mathcal{T}( {\tilde{\bm \xi }} )]^T|_{\tilde{\bm \xi } =\hat{\bm \xi }}.
\end{equation}

\section{  }\label{sect:appendixB}
\noindent\textbf{ \underline{Derivation of $q(\bm \theta)$ in (\ref{VIupdat2})}}

Since $\mathbb{E}_{  q(\bm s)}
 \left\{\ln  p(\bm \theta) \right\}=\ln p(\bm \theta)$, we have $\exp \left\{\mathbb{E}_{  q(\bm s)}
 \left\{\ln  p(\bm \theta) \right\} \right\}= p(\bm \theta)$.
Moreover, as
$\exp \left\{\mathbb{E}_{  q(\bm s)}
 \left\{\ln   p(\bm s)\right\}\right\}$ is a constant, (\ref{VIupdat2}) can be  simplified as
\begin{equation} \label{VIupdat1Appendix}
\begin{split}
q(\bm \theta)
\propto
p(\bm \theta)\exp \left\{\mathbb{E}_{  q(\bm s)}
 \left\{\ln   p(\bm z|\bm \theta, \bm s)\right\}\right\}.
\end{split}
\end{equation}
Next, we   perform the computation of $\exp \left\{\mathbb{E}_{  q(\bm s)}
 \left\{\ln   p(\bm z|\bm \theta, \bm s)\right\}\right\}$.
According to (\ref{linear}) and (\ref{linearPMU}), we have
\begin{equation}\label{51}
\begin{split}
&\exp \bigg\{\mathbb{E}_{  q(\bm s)}
 \big\{\ln   p(\bm z|\bm \theta, \bm s)\big\}\bigg\}\\
\propto & \exp \bigg\{\mathbb{E}_{  q(\bm s)}
 \big\{\sum_{i\in \mathcal{P}}
-\frac{\sigma_i^{-2}}{2}||\bm z_i-(\sum_{j\in  \mathcal{M}(i)}  \bm H_{ij}\bm s_j
+
\theta_i\sum_{j\in  \mathcal{M}(i)}   \bm G_{ij}\bm s_j)||^2\big\}\bigg\}.
\end{split}
\end{equation}
By expanding the squared norm and dropping the terms   irrelevant to $\theta_i$, (\ref{51}) is simpified as
\begin{equation}\label{52}
\begin{split}
&\exp \bigg\{\mathbb{E}_{  q(\bm s)}
 \left\{\ln   p(\bm z|\bm \theta, \bm s)\right\}\bigg\}\\
\propto &
\prod_{i\in \mathcal{P}}
\exp\Bigg\{
 - \frac{\sigma_i^{-2}}{2}\Big[-2\theta_i\textrm{Tr}\{\bm z_i\sum_{j\in \mathcal{M}(i)}
 (\bm G_{ij}\bm \mu_{\bm s_j})^T - \bm A_{i,1}\}  + \theta_i^2\textrm{Tr}\{\bm A_{i,2}\}
 \Big] \Bigg\}\\
\propto &
\prod_{i\in \mathcal{P}}
\mathcal{N}(\theta_i|\textrm{Tr}\{\bm z_i\sum_{j\in \mathcal{M}(i)}
 (\bm G_{ij}\bm \mu_{\bm s_j})^T - \bm A_{i,1}\}/\textrm{Tr}\{\bm A_{i,2}\},\sigma_i^2/\textrm{Tr}\{\bm A_{i,2}\}),
\end{split}
\end{equation}
where the last line comes from completing the square on the term inside the exponential and
${\bm A}_{i,1}
=\sum_{j\in  \mathcal{M}(i)}\bm H_{ij}\big({\bm P}_{j,j} + {\bm \mu}_j {\bm\mu}_j^T\big)\bm G^T_{ij}
+
\sum_{j,k\in  \mathcal{M}(i), j\neq k}\bm H_{ij}\big({\bm P}_{j,k} + {\bm \mu}_j {\bm\mu}_k^T\big)\bm G^T_{ik} $
and
${\bm A}_{i,2}
=
\sum_{j\in  \mathcal{M}(i)}\bm G_{ij}\big({\bm P}_{j,j} + {\bm \mu}_j {\bm\mu}_j^T\big)\bm G^T_{ij}
+
\sum_{j,k\in  \mathcal{M}(i), j\neq k}\bm G_{ij}\big({\bm P}_{j,k} + {\bm \mu}_j {\bm\mu}_k^T\big)\bm G^T_{ik} $.

Substituting $p(\bm\theta)$ from (\ref{thetaPrior}) and $\exp \big\{\mathbb{E}_{  q(\bm s)}
 \left\{\ln   p(\bm z|\bm \theta, \bm s)\right\}\big\}$ from (\ref{52})  into (\ref{VIupdat1Appendix}), we obtain
\begin{equation}\label{be-theta2-appen}
\begin{split}
q(\bm \theta)
\propto&
\prod_{i\in \mathcal{P}}
\frac{U(\theta_i - \underline{\theta_i} )-U(\theta_i - \bar{\theta}_{i} )}
{\textrm{erf}(\frac{\bar{\theta}_{i}-\tilde{v}_i}{\Tilde{C}^{1/2}_i})
     -\textrm{erf}(\frac{\underline{\theta_i}-\tilde{v}_i}{\Tilde{C}^{1/2}_i})}
\mathcal{N}(\theta_i|\tilde{v}_i, \Tilde{C}_i)\\
&\times \mathcal{N}(\theta_i|\textrm{Tr}\{\bm z_i\sum_{j\in \mathcal{M}(i)}
 (\bm G_{ij}\bm \mu_{\bm s_j})^T - \bm A_{i,1}\}/\textrm{Tr}\{\bm A_{i,2}\},\sigma_i^2/\textrm{Tr}\{\bm A_{i,2}\})   \\
\propto&
\prod_{i\in \mathcal{P}}
\frac{U(\theta_i - \underline{\theta_i} )-U(\theta_i - \bar{\theta}_{i} )}
{\textrm{erf}(\frac{\bar{\theta}_{i}- {v}_i}{ {C}^{1/2}_i})
     -\textrm{erf}(\frac{\underline{\theta_i}- {v}_i}{ {C}^{1/2}_i})}
\mathcal{N}(\theta_i| {v}_i,  {C}_i)
\end{split}
\end{equation}
with
\begin{equation}\label{truncatedC-appen}
{C}_i
= \frac{\tilde{C}_i }{\sigma_i^{-2}\textrm{Tr}\{
{\bm A}_{i,2}\}\tilde{C}_i+ 1 },
\end{equation}
\begin{equation}\label{truncatedv-appen}
{v}_i
={C}_i\big[\tilde{v}_i/\tilde{C}_i+ \sigma_i^{-2}
\textrm{Tr}\big\{\bm z_i\sum_{j\in  \mathcal{M}(i)}
(\bm G_{ij}{\bm \mu}_j)^T
-{\bm A}_{i,1}
\big\}\big].
\end{equation}
It is recognized that (\ref{be-theta2-appen}) is in the form of a truncated Gaussian pdf.  That is, $q(\bm \theta) \propto
\prod_{i\in \mathcal{P}}\mathcal{TN}(\theta_i|\underline{\theta_i},\bar{\theta}_{i},
{v}_i, {C}_i)$.

\noindent\textbf{ \underline{Derivation of $q(\bm s)$ in (\ref{VIupdat3})} }

Similar to the arguments for arriving at (\ref{VIupdat1Appendix}), (\ref{VIupdat3}) can be simplified as
\begin{equation} \label{VIupdatAppendix}
\begin{split}
q(\bm s)
\propto
p(\bm s)\exp \left\{\mathbb{E}_{ q(\bm\theta) }
\left\{\ln  p(\bm z|\bm \theta, \bm s)\right\}\right\}.
\end{split}
\end{equation}
For $\exp \left\{\mathbb{E}_{ q(\bm\theta) }
\left\{\ln  p(\bm z|\bm \theta, \bm s)\right\}\right\}$, it can be computed as
\begin{equation} \label{sappendix}
\begin{split}
&\exp \left\{\mathbb{E}_{ q(\bm\theta) }
\left\{\ln  p(\bm z|\bm \theta, \bm s)\right\}\right\}\\
\propto & \exp \bigg\{\mathbb{E}_{  q(\bm \theta)}
 \big\{-\frac{1}{2}||\bm z - (\bm H  +
\bm\Theta\bm G)\bm s||_{\bm R^{-1}}^2\big\}\bigg\} \\
=& \exp \bigg\{
 -\frac{1}{2} \big[\bm z^T\bm R^{-1}\bm z - 2\bm z^T\bm R^{-1}(\bm H  +
\mathbb{E}_{ q(\bm\theta) }\{\bm\Theta\}\bm G)\bm s \\
&
\qquad \quad + \bm s^T\bm H^T\bm R^{-1}\bm H\bm s
+2\bm s^T\bm H^T\bm R^{-1}\mathbb{E}_{ q(\bm\theta) }\{\bm\Theta\}\bm G\bm s \\
&\qquad \quad +\bm s^T\bm G^T\bm R^{-1}\mathbb{E}_{ q(\bm\theta) }\{\bm\Theta^2\}\bm G\bm s
\big]\bigg\}\\
=& \exp \bigg\{
 -\frac{1}{2} [\bm z-(\bm H  +
\bm \Omega\bm G)\bm s]^T\bm R^{-1} [\bm z-(\bm H  +
\bm \Omega\bm G)\bm s]
 -\frac{1}{2}  \bm s^T\bm G^T(\bm \Lambda - \bm \Omega^2)\bm R^{-1}\bm G\bm s
  \bigg\} \\
\propto &
\mathcal{N}\big(\bm z|(\bm H  + {\bm \Omega} \bm G)\bm s,
\bm R \big)
\times
\mathcal{N}\big(\bm s|\bm 0, (\bm G^T(\bm \Lambda - \bm \Omega^2)\bm R^{-1}\bm G)^{-1}\big)
\end{split}
\end{equation}
where ${\bm \Omega}=\mathbb{E}_{ q(\bm\theta) }\{\bm\Theta\}  \triangleq \textrm{Bldiag}\{{\varpi}_i\bm I_{2|\mathcal{M}(i)|}, \ldots, {\varpi}_j\bm I_{2|\mathcal{M}(j)|}\}$
and
${\bm \Lambda}=\mathbb{E}_{ q(\bm\theta) }\{\bm\Theta^2\}\triangleq \textrm{Bldiag}\{{\tau}_{i}\bm I_{2|\mathcal{M}(i)|},\\ \ldots, {\tau}_{j}\bm I_{2|\mathcal{M}(j)|}\}$.

By substituting the prior distribution
$p(\bm s)$ from
(\ref{sPrior}), and $\exp \left\{\mathbb{E}_{ q(\bm\theta) }
\left\{\ln  p(\bm z|\bm \theta, \bm s)\right\}\right\}$ from (\ref{sappendix})
 into (\ref{VIupdatAppendix}), and after some algebraic manipulations \cite[pp. 326]{Kay}, we obtain
\begin{equation}\label{be-s2-appexd}
\begin{split}
q(\bm s)
\propto
\mathcal{N}\big(\bm s|{\bm \mu} , {\bm P}\big)
\end{split}
\end{equation}
with the  covariance and mean given by
\begin{equation}\label{s-meanupdated-appexd}
\begin{split}
{\bm \mu}
=&
\bm \Gamma_{\textrm{SCADA}}^{-1} \hat{\bm s}_{\textrm{SCADA}} + \bm\Upsilon(\bm H  +  {\bm \Omega} \bm G)^T
\big[(\bm H  +  {\bm \Omega} \bm G){\bm\Upsilon} (\bm H  +  {\bm \Omega} \bm G)^T + \bm R \big]^{-1}\\
&\times[\bm z -(\bm H  +  {\bm \Omega} \bm G)
\bm \Gamma_{\textrm{SCADA}}^{-1} \hat{\bm s}_{\textrm{SCADA}}],
\end{split}
\end{equation}
\begin{equation}\label{s-Covupdated-appexd}
{\bm P}
={\bm\Upsilon}-{\bm\Upsilon}(\bm H  +  {\bm \Omega} \bm G)^T
\big[(\bm H  +  {\bm \Omega} \bm G){\bm\Upsilon} (\bm H  +  {\bm \Omega} \bm G)^T + \bm R \big]^{-1}
(\bm H  +  {\bm \Omega} \bm G){\bm\Upsilon},
\end{equation}
respectively, where
${\bm\Upsilon} = [\bm \Gamma_{\textrm{SCADA}}^{-1} + (\bm G^T
({\bm \Lambda}- {\bm \Omega}^2)\bm R^{-1}
\bm G)^{-1}]^{-1}.
$


\section{ }\label{appendixC}

From (\ref{linear}), it can be obtained that
$p(\bm z_j|\theta_j, \{\bm s_{\tilde{k}}\}_{\tilde{k}\in \mathcal{M}(j)})\propto
\exp\big\{-\frac{\sigma_j^2}{2}||\bm z_j-(\sum_{\tilde{k}\in  \mathcal{M}(j)}\bm H_{j\tilde{k}}\bm s_{\tilde{k}}
+
\theta_j\sum_{\tilde{k}\in  \mathcal{M}(j)}   \bm G_{j\tilde{k}}\bm s_{\tilde{k}})||^2\big\}$.
By expanding the squared norm and dropping the terms irrelevant to $\bm s_i$, we have
\begin{equation}
\begin{split}
&\ln p(\bm z_j|\theta_j, \{\bm s_{\tilde{k}}\}_{\tilde{k}\in \mathcal{M}(j)})
\\
\propto &
-\frac{\sigma_j^{-2}}{2}
\bigg[
-2\bm z_j^T(\bm H_{ji}+ \theta_j\bm G_{ji})\bm s_i \\
&+2
 \sum_{k\in \mathcal{M}(j)\setminus i}
\bm s_k^T\big\{(\bm H_{ji}+ \theta_j\bm G_{ji})^T(\bm H_{jk}+ \theta_j\bm G_{jk})\bm s_i \big\}   \\
&+\bm s_i^T(\bm H_{ji}+ \theta_j\bm G_{ji})^T(\bm H_{ji}+ \theta_j\bm G_{ji})\bm s_i
\bigg].
\end{split}
\end{equation}
Taking expectation with respect to $\theta_j$ and $\{\bm s_k\}_{k\in \mathcal{M}(j)\setminus i}$ over the above equation, we have
\begin{equation}\label{s-completesquare}
\begin{split}
&\exp \left\{\mathbb{E}_{b(\theta_j)\prod_{k\in \mathcal{M}(j)\setminus i}b(\bm s_k)}
\{  \ln p(\bm z_j|\theta_j, \{\bm s_{\tilde{k}}\}_{\tilde{k}\in \mathcal{M}(j)})\}\right\}\\
\propto&
\exp \bigg\{
-\frac{\sigma_j^{-2}}{2}
\bigg[
-2\bm z_j^T(\bm H_{ji}+ \varpi_j\bm G_{ji})\bm s_i \\
&+2
 \sum_{k\in \mathcal{M}(j)\setminus i}
\bm \mu^T_k\big\{\bm H_{ji}^T\bm H_{jk}+ \varpi_j(\bm G_{ji}^T\bm H_{jk}+ \bm H_{ji}^T\bm G_{jk})+
\tau_j\bm G_{ji}^T\bm G_{jk}  \big\} \bm s_i  \\
&+\bm s_i^T\{\bm H_{ji}^T\bm H_{ji} + \varpi_j(\bm G_{ji}^T\bm H_{ji}+\bm H_{ji}^T\bm G_{ji})+ \tau_{j}\bm G_{ji}^T\bm G_{ji}\}\bm s_i
\bigg]
\bigg\}.
\end{split}
\end{equation}
Then,  completing the  square for the term $\bm s_i$ in (\ref{s-completesquare}) leads to
\begin{equation}
m_{j\rightarrow i}(\bm s_i)\propto \mathcal{N}(\bm s_i|\bm v_{j\rightarrow i}, \bm C_{j\rightarrow i}),
\end{equation}
with
\begin{equation}\label{vmp12}
\bm C_{j\rightarrow i}=\sigma_j^2\big[\bm H_{ji}^T\bm H_{ji} + \varpi_j(\bm G_{ji}^T\bm H_{ji}+\bm H_{ji}^T\bm G_{ji})+ \tau_{j}\bm G_{ji}^T\bm G_{ji}\big]^{-1},
\end{equation}
\begin{equation}\label{vmp22}
\begin{split}
\bm v_{j\rightarrow i}
=&\sigma_j^{-2}\bm C_{j\rightarrow i}\bigg\{
(\bm H_{ji}+ \varpi_j\bm G_{ji})^T\bm z_j\\
&
-
\sum_{k\in \mathcal{M}(j)\setminus i}
\big[\bm H_{ji}^T\bm H_{jk}+ \varpi_j(\bm G_{ji}^T\bm H_{jk}+ \bm H_{ji}^T\bm G_{jk})+
\tau_j\bm G_{ji}^T\bm G_{jk}  \big]^T\bm \mu_k  \bigg\}.
\end{split}
\end{equation}

\bibliographystyle{IEEEtran}

\newpage

\end{document}